\documentclass[12pt, prd, preprint, amsmath, amssymb, aps, superscriptaddress,eqsecnum, floats,nofootinbib,preprintnumbers]{revtex4-1}
\usepackage{color,graphicx,epsfig}
\usepackage{ mathrsfs }
\usepackage{ifpdf}
\usepackage{amsmath}
\usepackage{bm}
\usepackage{color}
\usepackage[dvipsnames]{xcolor}
\usepackage[english]{babel}
\usepackage{graphicx}%
\usepackage{amsfonts}%
\usepackage{amssymb}
\usepackage{braket}
\usepackage{rotating}
\usepackage{array}
\usepackage[colorlinks=true,citecolor=blue,urlcolor=blue,linktocpage=true, linkcolor=blue]{hyperref}
\usepackage{multirow}
\usepackage{marvosym}
\usepackage{cancel}
\usepackage{enumitem}
\setlist{nosep}

\definecolor{nicered}{rgb}{0.7,0.1,0.1}
\definecolor{nicegreen}{rgb}{0.1,0.5,0.1}
\hypersetup{colorlinks,citecolor= nicegreen,linkcolor= nicered}

\newenvironment{Eqnarray}{\arraycolsep 0.14em\begin{eqnarray}}{\end{eqnarray}}
\def\beqa{\begin{Eqnarray}}
\def\eeqa{\end{Eqnarray}}
\newcommand{\no}{\nonumber}
\newcommand{\abs}[1]{\lvert#1\rvert}

\newcommand{\br}{{\rm BR}}
\newcommand{\sm}{{\rm SM}}

\newcommand{\pc}{priv. comb.}
\newcommand{\unit}[1]{\,\ensuremath{\textrm{#1}}}

\newcommand{\mev}{\unit{MeV}}
\newcommand{\gev}{\unit{GeV}}
\newcommand{\tev}{\unit{TeV}}

\newcommand{\ifb}{\unit{fb}^{-1}}

\newcommand{\cl}{\unit{C.\,L.}}

\newcommand{\beq}{\begin{equation}}
\newcommand{\eeq}{\end{equation}}
\newcommand{\bea}{\begin{eqnarray}}
\newcommand{\eea}{\end{eqnarray}}

\newcommand{\be}{\begin{equation}}
\newcommand{\ee}{\end{equation}}
\def\lsim{\mathrel{\rlap{\lower4pt\hbox{\hskip1pt$\sim$}}
    \raise1pt\hbox{$<$}}}         
\def\gsim{\mathrel{\rlap{\lower4pt\hbox{\hskip1pt$\sim$}}
    \raise1pt\hbox{$>$}}}         
\begin{document}
\fontsize{11}{15}\selectfont
\title{
\hfill\small{\textnormal{FERMILAB-PUB-20-072-T, EFI-20-3}}\\ \vspace*{0.5cm}
\large{
$CP$ violation from $\tau$, $t$ and $b$ dimension-6 Yukawa couplings \\
-- interplay of baryogenesis, EDM and Higgs physics}}

\author{Elina Fuchs}
\email{elinafuchs@uchicago.edu}
\affiliation{Department of Particle Physics and Astrophysics, Weizmann Institute of Science, Rehovot, Israel 7610001}
\affiliation{Fermilab, Theory Department, Batavia, IL 60510, USA}
\affiliation{University of Chicago, Department of Physics, Chicago, IL 60637, USA}

\author{Marta Losada}
\email{marta.losada@nyu.edu}
\affiliation{New York University Abu Dhabi, PO Box 129188, Saadiyat Island, Abu Dhabi, United Arab Emirates}

\author{Yosef Nir}
\email{yosef.nir@weizmann.ac.il}
\affiliation{Department of Particle Physics and Astrophysics, Weizmann Institute of Science, Rehovot, Israel 7610001}

\author{Yehonatan Viernik}
\email{yehonatan.viernik@weizmann.ac.il}
\affiliation{Department of Particle Physics and Astrophysics, Weizmann Institute of Science, Rehovot, Israel 7610001}

\begin{abstract}
\noindent
\fontsize{12}{13}\selectfont
We explore the implications of the Standard Model effective field theory (SMEFT) with dimension-six terms involving the Higgs boson and third-generation fermion fields on the rate of Higgs boson production and decay into fermions, on the electric dipole moments (EDMs) of the electron, and on the baryon asymmetry of the Universe. We study the consequences of allowing these additional terms for each flavor separately and for combinations of two flavors.
We find that a complex $\tau$ Yukawa coupling can account for the observed baryon asymmetry $Y_B^{\rm obs}$ within current LHC and EDM bounds. A complex $b$ ($t$) Yukawa coupling can account for $4\%$ ($2\%$) of $Y_B^{\rm obs}$, whereas a combination of the two can reach $12\%$. Combining $\tau$ with either $t$ or $b$ enlarges the viable parameter space owing to cancellations in the EDM and in either Higgs production times decay or the total Higgs width, respectively. Interestingly, in such a scenario there exists a region in parameter space where the SMEFT contributions to the electron EDM cancel and collider signal strengths are precisely SM-like, while producing sufficient baryon asymmetry. Measuring $CP$ violation in Higgs decays to $\tau$ leptons is the smoking gun for this scenario.
\end{abstract}
\maketitle
\thispagestyle{empty} 
\fontsize{12}{9.5}\selectfont
\tableofcontents
\thispagestyle{empty} 
\fontsize{12}{13}\selectfont

\clearpage
\section{Introduction}
\setcounter{page}{1}
The first decade of the LHC experiments led to significant progress in our understanding of Nature. Two very important aspects of this progress have been the following:
\begin{itemize}
\item A new scalar particle has been discovered~\cite{Aad:2012tfa,Chatrchyan:2012xdj} with properties that fit, within present experimental accuracy, to those of the Standard Model (SM) Higgs boson \cite{Khachatryan:2016vau,Aad:2019mbh,Sirunyan:2018koj,deFlorian:2016spz,Bechtle:2014ewa,Ellis:2018gqa}.
\item No other new elementary particles have been discovered, with lower bounds on the mass of large classes of such hypothetical particles at the TeV scale~\cite{Sirunyan:2019xwh,Aad:2019hjw}.
\end{itemize}
This situation makes it plausible that the scale of new physics is high enough above the electroweak scale that its effects can be parameterized via higher-dimension operators, and motivates an interpretation of experimental results in the framework of the Standard Model effective field theory (SMEFT). The Higgs program provides a unique window into various classes of such higher-dimension terms in the Lagrangian.

We are particularly interested in dimension-six operators~\cite{Buchmuller:1982ye,Grzadkowski:2010es} that couple the Higgs boson field to fermion fields. The presence of these additional terms provides two important features: novel $CP$-violating interactions~\cite{deVries:2017ncy} and violation of the SM relation between the fermion mass and its Yukawa coupling. These features lead to interesting consequences:
\begin{itemize}
\item Modifications to the Higgs production and decay rates, which can potentially be discovered by collider experiments;
\item New contributions to the electric dipole moment of the electron (EDM), potentially within present or near-future reach of experiments;
\item New contributions to the baryon asymmetry via electroweak baryogenesis (EWBG), with the potential of opening a window to solving this long-standing problem.
\end{itemize}
In this work we study these effects and their interplay for the fermions of the third generation, $\tau$, $b$ and $t$. The details and implications for the muon can be found in Ref.~\cite{Fuchs:2019ore}. For previous work on the possible role of third-generation fermions in the aspect of $CP$ violation for EWBG, see {\it e.g.} Refs.~\cite{deVries:2017ncy,Kobakhidze:2015xlz} for quarks, and Refs.~\cite{Guo:2016ixx,deVries:2018tgs} for the tau-lepton.

This paper is organized as follows. In Section \ref{sec:framework} we introduce our theoretical framework and a useful parameterization of the coupling constants. Sections \ref{sec:ewb} and \ref{sec:edm} present the effect of the dimension-six terms on electroweak baryogenesis and the electron EDM, respectively. We then focus in Section \ref{sec:collider} on the LHC results of Higgs boson decay rates to fermion pairs and vector boson pairs from various production channels, as well as on the modifications of the total Higgs width. We derive the constraints these measurements impose on the new physics captured in the dimension-six terms. Sections \ref{sec:results1} and \ref{sec:results2} present our results for single species and for combinations of two fermions. In Section \ref{sec:conclusion} we summarize and present our conclusions.

\section{Dimension-six complex Yukawa terms}
\label{sec:framework}
We consider the following dimension-four and dimension-six Yukawa-type Lagrangian terms for the third-generation fermions (similarly to Refs.~\cite{deVries:2017ncy,deVries:2018tgs}, but allowing also for a real part of the dimension-six term):
\beq\label{eq:lagd4}
{\cal L}_{\rm Yuk}= y_f\overline{F_L}F_RH+
\frac{1}{\Lambda^2}(X_R^f+iX_I^f)|H|^2\overline{F_L}F_RH + {\rm h.c.}
\eeq
Here $F_L$ is the $SU(2)$-doublet field containing $F=t,b,\tau$, $F_R$ is the corresponding $SU(2)$-singlet field, $H$ is the Higgs doublet field, and $\Lambda$ is the mass scale of new physics. Without loss of generality, we take $y_f$ to be real. Substituting in the unitary gauge
\beq
H=\frac{1}{\sqrt2}(v+h),
\eeq
leads to the following mass term and $h$-Yukawa couplings:
\beqa\label{eq:yukd46}
{\cal L}_{f}&=&
\frac{y_f v}{\sqrt2}
\left[1+\frac{v^2}{2\Lambda^2}\frac{X_R^f+iX_I^f}{y_f}\right]\overline{f_L}f_R
+\frac{y_f}{\sqrt2}\left[1+\frac{3v^2}{2\Lambda^2}\frac{X_R^f+iX_I^f}{y_f}\right]\overline{f_L}f_R h\no\\
&+&\frac{3v}{2\sqrt{2}\Lambda^2}(X_R^f+iX_I^f)\overline{f_L}f_R hh
+\frac{1}{2\sqrt{2}\Lambda^2}(X_R^f+iX_I^f)\overline{f_L}f_R hhh.\
\eeqa
We define the ratio of the dim-6 to the dim-4 contribution to a fermion mass as our useful coordinates to be used in the following:
\beq
T_R^f\equiv\frac{v^2}{2\Lambda^2}\frac{X_R^f}{y_f},\ \ \
T_I^f\equiv\frac{v^2}{2\Lambda^2}\frac{X_I^f}{y_f}.
\eeq
Thus the coefficients of the mass and Yukawa terms in Eq. (\ref{eq:yukd46}) have the following values:
\beq
\label{eq:m_lambda}
m_f=\frac{y_fv}{\sqrt2}\left(1+T_R^f+iT_I^f\right),\ \ \
\lambda_f=\frac{y_f}{\sqrt2}\left(1+3T_R^f+3iT_I^f\right)\,.
\eeq

Once we add the dimension-six terms, we are no longer in the basis of real fermion masses. To have $m_f$ real in the $m_f\overline{f_L}f_R$ term, we transform $f_R\to e^{i\theta_f}f_R$ by $\theta_f$ which satisfies
\beq
\tan\theta_f=\frac{T_I^f}{1+T_R^f}.
\eeq
Then, in the mass basis with a real value for the mass,
\beqa\label{eq:mfytrti}
m_f&=&\frac{y_fv}{\sqrt2}\sqrt{(1+T_R^f)^2+T_I^{f2}},
\eeqa
we have the following Yukawa coupling:
\beqa\label{eq:massyuk}
\lambda_f&=&
\frac{y_f}{\sqrt2}\frac{1+4T_R^f+3T_R^{f2}+3T_I^{f2}+2iT_I^f}{\sqrt{(1+T_R^f)^2+T_I^{f2}}}.
\eeqa

The dim-4 coupling $ y_f $ can be written in terms of $ T_R^f,T_I^f $ via the expression \eqref{eq:mfytrti} for the mass. In turn, this can be related to the SM Yukawa coupling via
\begin{align}\label{eq:yfyfSM}
\left(\frac{y_f}{y_f^\sm}\right)^2&=\frac{1}{(1+T^f_R)^2+T_I^{f2}}\,.
\end{align}
Thus, the full setup of Eq.~(\ref{eq:lagd4}) is described by two free parameters per fermion, $T_R^f$ and $T_I^f$.

\section{The baryon asymmetry $Y_B$}
\label{sec:ewb}
The value of the baryon asymmetry is extracted from CMB measurements. It is given by
$\Omega_bh^2=0.02226(23)$~\cite{Tanabashi:2018oca} or, equivalently,
\beq\label{eq:ybobs}
Y_B^{\rm obs}=(8.59\pm0.08)\times10^{-11}.
\eeq

Electroweak baryogenesis is the mechanism through which a non-zero value of the baryon number density is obtained during the electroweak phase transition (EWPT). As bubbles expand to fill the universe with the non-zero vacuum expectation value of the Higgs field, $CP$-violating interactions across the bubble wall create a chiral asymmetry, which is then converted to a baryon asymmetry by the weak sphaleron process. Electroweak baryogenesis requires two ingredients that go beyond the SM:
\begin{itemize}
\item New sources of $CP$ violation~\cite{Gavela:1993ts,Huet:1994jb};
\item Modification of the EWPT such that it is strongly first order (rather than a smooth crossover, as is the case in the SM~\cite{Kajantie:1996mn,Csikor:1998eu}).
\end{itemize}
In this work we focus on the aspect of $CP$ violation. We thus make the following assumptions regarding the EWPT:
\begin{itemize}
\item There are additional degrees of freedom that lead to a strongly first-order EWPT;
\item These additional degrees of freedom do not significantly affect the interactions of the SM fermion fields across the expanding bubble wall;
\item There are no additional sources of $CP$ violation from the interactions of these degrees of freedom that would significantly modify the resulting value of the baryon asymmetry.
\end{itemize}

\subsection{Particle dynamics}
We calculate the final matter-antimatter asymmetry in the Closed Time Path formalism, following \cite{Lee:2004we,White:2016nbo,deVries:2017ncy,deVries:2018tgs}. For simplicity, we provisionally consider the case that only one active fermion species with a non-zero dimension-six term provides a source for generating the asymmetry. Here we neglect first and second lepton generations due to the smallness of their Yukawa couplings (for muon-driven EWBG see Ref.~\cite{Fuchs:2019ore}). Interactions of light quarks are neglected as well, but they participate in the strong sphaleron process, which is fast at high temperatures. Gauge interactions are fast enough to be considered in equilibrium. The dimension-six term leads to both $CP$-odd and $CP$-even processes that compete to produce and wash out a $CP$ asymmetry.
The summary of the process is as follows:
{\footnotesize }
\begin{itemize}
\item $CP$-violating interactions across the expanding bubble wall generate a chiral asymmetry, while $CP$-conserving interactions wash out the generated asymmetry.
\item The strong sphaleron process produces further washout in the quark sector.
\item Some of the remaining asymmetry diffuses into the symmetric phase. Diffusion is dominantly affected by gauge interactions, hence it is more efficient for leptons than for quarks.
\item The weak sphaleron process is efficient only in the symmetric phase, acting on left-handed multiplets and changing baryon number.
\item The chemical potential due to the chiral asymmetry induces a preferred direction for the weak sphaleron, thus generating a baryon asymmetry.
\item Finally, the bubble wall catches up and freezes in the resulting baryon number density in the broken phase.
\end{itemize}

The full dynamics described above is encoded in a coupled set of differential equations, the transport equations, one for each flavor $f$:
\beq
\partial_\mu f^\mu = -\Gamma_M^f \mu_M^f-\Gamma_Y^f \mu_Y^f + \Gamma_{\text{ss}}^f \mu_{\text{ss}}  -\Gamma_{\text{ws}}^f \mu_{\text{ws}}^f+ S_f,
\eeq
where the relaxation and Yukawa rates $\Gamma_M^f,\Gamma_Y^f$ relate to $CP$-conserving interactions; the strong sphaleron rate $\Gamma_{\text{ss}}$ is non-zero only for quarks; and the weak sphaleron rate $\Gamma_{\text{ws}}^f$ is non-zero only for left-handed fermions. A chemical potential $ \mu $ is associated with each of these processes, and $ S_f $ is the $CP$-violating source, which does not admit a chemical potential.
The method we used to solve this set of equations and other details are presented in \cite{flnv}.

\subsection{Impact of $T_R$ and $T_I$}
The baryon asymmetry is proportional to the source, $Y_B \propto S\propto T_I$ at lowest order. Hence the $T_R$ dependence enters only from second order in $T_{R,I}\propto 1/\Lambda^2$ onward, thus at $\mathcal{O}(1/\Lambda^4)$.\footnote{At $\mathcal{O}(1/\Lambda^4)$, there are contributions from both dim-6 terms squared and dim-4 times dim-8 terms. The contribution from the former is, however, enhanced by $1/y_f$ compared to the latter, and therefore, for the $b$-quark and for the $\tau$-lepton, the latter can be neglected. For the effects of dim-8 terms on $Y_B$ from a complex $t$-Yukawa, see Ref.~\cite{deVries:2017ncy}.}

In addition to the dependence of the $CP$ violation source $S_f$ on the dimension-six terms,
\beq
S_f\propto{\cal I}m(m_f^* m_f^\prime)\propto y_f^2T_I^f,
\eeq
the relaxation rate $\Gamma_M$, which originates from two mass insertions, and the Yukawa rate $\Gamma_Y$, which originates from two Yukawa insertions, are rescaled by $T_R^f,T_I^f$-dependent factors:
\begin{align}
 \Gamma_M &\rightarrow \left[\frac{(1+r_{N0}^2T_R^f)^2+r_{N0}^2T_I^{f2}}{(1+T_R^f)^2+T_I^{f2}}\right] \Gamma_M\,,\no\\
 \Gamma_Y &\rightarrow \left[\frac{(1+3 r_{N0}^2T_R^f)^2+ (3r_{N0}^2T_I^f)^2}{(1+T_R^f)^2+T_I^{f2}}\right] \Gamma_Y\,,
\end{align}
Here $ r_{N0} \equiv v(T=T_N)/v(T=0) $, where $T_N$ is the nucleation temperature.

Our numerical calculation yields, to leading order\footnote{For large $T_I, T_R$, the dependence via Eq.~(\ref{eq:yfyfSM}) will become relevant.} in $T_I,T_R$, for $T_R=0$ and parameters as in Appendix~\ref{sec: benchmark} and \cite{flnv},
\beq\label{eq:YBttaub}
Y_B=8.6\times10^{-11}\times(51 T_I^t - 23 T_I^\tau - 0.44 T_I^b).
\eeq
We learn that the relevant range for each of the third generation fermions to account for the baryon asymmetry is
\beq
|T_I^t|={\cal O}(0.02),\ \ \ |T_I^\tau|={\cal O}(0.04),\ \ \ |T_I^b|>1.
\eeq
%

\section{The electron EDM $d_e$}
\label{sec:edm}
An upper bound on the electric dipole moment of the electron $|d_e|$ was recently obtained by the ACME collaboration \cite{Andreev:2018ayy}:
\beq\label{eq:deexp}
|d_e^{\rm max}|=1.1\times10^{-29}\ e\ {\rm cm} \text{~at~} 90\% \cl\,.
\eeq

The dimension-six terms contribute also to $d_e$. We rewrite the relevant results obtained in Ref. \cite{Panico:2018hal} in terms of our parameterization. The finite contributions from the Barr-Zee diagrams are given by
\beqa\label{eq:def}
\frac{d_e^{(t)}}{e}&\simeq&-\frac{16}{3}\frac{e^2}{(16\pi^2)^2}\frac{m_e}{m_t}\frac{v}{\Lambda^2}X_I^t
\left(2+\ln\frac{m_t^2}{m_h^2}\right),\\
\frac{d_e^{(b)}}{e}&\simeq&-4N_c Q_b^2\frac{e^2}{(16\pi^2)^2}\frac{m_em_b}{m_h^2}\frac{v}{\Lambda^2}X_I^b
\left(\frac{\pi^2}{3}+\ln^2\frac{m_b^2}{m_h^2}\right),\no\\
\frac{d_e^{(\tau)}}{e}&\simeq&-4 Q_\tau^2\frac{e^2}{(16\pi^2)^2}\frac{m_em_\tau}{m_h^2}\frac{v}{\Lambda^2}X_I^{\tau}
\left(\frac{\pi^2}{3}+\ln^2\frac{m_\tau^2}{m_h^2}\right).\no
\eeqa
Working in the real mass basis and using the full Yukawa interaction \eqref{eq:massyuk}, the sum of the $t,b,\tau$ finite contributions becomes
\beqa\label{eq:d_e expression}
\frac{d_e}{e}&\simeq&-\frac{32\sqrt{2}}{3}\frac{e^2}{(16\pi^2)^2}\frac{m_e}{v^2}
\left[\left(2+\ln\frac{m_t^2}{m_h^2}\right)\left( \frac{y_t}{y_t^{\sm}} \right)^2 T_I^t \right.\\
&&\left.+\frac{1}{4}\left(\frac{\pi^2}{3}+\ln^2\frac{m_b^2}{m_h^2}\right)\frac{m_b^2}{m_h^2} \left( \frac{y_b}{y_b^{\sm}} \right)^2 T_I^b
+\frac{3}{4}\left(\frac{\pi^2}{3}+\ln^2\frac{m_\tau^2}{m_h^2}\right)\frac{m_\tau^2}{m_h^2}
\left( \frac{y_\tau }{y_\tau ^{\sm}} \right)^2 T_I^{\tau}\right].\no
\eeqa
Hence, the leading dim-6 dependence is on $T_I^f$, but through $(y_f/y_f^\sm)$ given in Eq.~(\ref{eq:yfyfSM}) also on $T_R^f$:
\begin{equation}\label{eq: d_e coefficients}
d_e \approx 1.1 \times 10^{-29} \;e\; \text{cm}\; \times
\left[2223 \left(\frac{y_t}{y_t^\sm}\right)^2 T_I^t
+ 9.6 \left(\frac{y_\tau}{y_\tau^\sm}\right)^2 T_I^\tau
+ 11.6 \left(\frac{y_b}{y_b^\sm}\right)^2 T_I^b \right].
\end{equation}
We learn that for $y_f={\cal O}(y_f^\sm)$, the sensitivity of the current searches for $d_e$ is
\beq
\label{eq:TIEDM}
T_I^t={\cal O}(0.0004),\ \ \ T_I^\tau={\cal O}(0.1),\ \ \ T_I^b={\cal O}(0.09).
\eeq

Additional constraints arise from measurements of the electric dipole moments of the neutron, mercury or thalium, see Refs.~\cite{Brod:2013cka,Cirigliano:2016nyn,Brod:2018pli} and references therein. However, both the hadronic and matrix element uncertainties, and possible cancellations \cite{Fuyuto:2019svr} from CP-odd contributions involving the top and/or the bottom quark to these observables via Barr-Zee diagrams, the Weinberg operator, chromo-electric dipole moments for light quarks {\it etc.}, make the constraints on $T_{I}^{t,b}$ weaker. In the case of a nonzero $T_I^{\tau}$, there is only a Barr-Zee type contribution to the neutron EDM. However, given the current experimental upper bound on the neutron EDM \cite{Abel:2020gbr}, it does not provide a stronger constraint on $T_I^{\tau}$.

\section{Higgs production and decay}
\label{sec:collider}
In this section, we derive the dependence of Higgs production and decay rates on $T_R$ and $T_I$ for cases where either one or two Yukawa couplings of third generation fermions are modified by dim-6 contributions. We then present those collider processes that are most sensitive to the considered coupling modifications, and their current experimental bound. In order to obtain the strongest available bound from Run-2 data, we combine the published values of ATLAS and CMS in a naive theorists' approach as the weighted mean with symmetric upper and lower uncertainties. The details of individual rates and their underlying data sets are summarized in Appendix~\ref{app:munumbers}.

\subsection{Signal strength}

The Higgs signal strength of production mode $I$ (such as $I=$ ggF) and decay channel into a final state $F$ is defined as
\begin{align}\label{eq:muffgeneral}
 \mu_I^F&\equiv
  \frac{\sigma_I(pp\to h)\,\cdot \Gamma(h\to F)/\Gamma_h}{[\sigma_I(pp\to h)\,\cdot\Gamma(h\to F)/\Gamma_h]_{\rm SM}},
\end{align}
where $\Gamma_h$ is the total Higgs width. To extract the dependence of $\mu_I^F$ on the SMEFT parameters, it is convenient to define the dimensionless parameters $r_f$:
\beq\label{eq:defrf}
r_f\equiv \frac{|\lambda_f|^2/|\lambda_f^\sm|^2}{|m_f|^2/|m_f^\sm|^2} =  \frac{(1+3T_R^f)^2 + 9T_I^{f2}}{(1+T_R^f)^2 + T_I^{f2}}\,.
\eeq

\subsubsection{Production rates}
The main production modes of the discovered Higgs boson at $m_h=125\gev$ are gluon fusion (ggF), associated $t\bar t h$ and $th$ production (together denoted as $tth$), vector-boson associated production ($Vh$, $V=Z,W$) and vector-boson fusion (VBF). The ggF and $tth$ production rates are proportional to $|\lambda_t|^2$ (neglecting the very small $b$ quark contribution), whereas the $Vh$ and VBF rates do not depend on any Yukawa coupling:
\beqa
\sigma_{\rm ggF}/\sigma_{\rm ggF}^{\rm SM}&=&\sigma_{tth}/\sigma_{tth}^{\rm SM}=r_t,\no\\
\sigma_{Vh}/\sigma_{Vh}^{\rm SM}&=&\sigma_{\rm VBF}/\sigma_{\rm VBF}^{\rm SM}=1.
\eeqa

\subsubsection{Decay rates}
We consider modifications of decays into fermion pairs $F=\bar f f$. We obtain:
\beq
\Gamma(h\to f\bar f)/[\Gamma(h\to f\bar f)]^{\rm SM}=r_f\ \ \ (f=b,\tau) .
\eeq
We also make use of decays into the weak vector-bosons:
\beq
\Gamma(h\to VV^*)/[\Gamma(h\to VV^*)]^{\rm SM}=1\ \ \ (V=W,Z) .
\eeq

\subsubsection{The total Higgs width}
Within our framework, the total Higgs width is affected by modification of $\lambda_b$, via $\Gamma(h\to b\bar b)$ with SM branching ratio $\br_b^{\rm SM}=0.58$, by modification of $\lambda_\tau$, via $\Gamma(h\to\tau^+\tau^-)$ with $\br_\tau^{\rm SM}\sim0.063$, and by modification of $\lambda_t$, via $\Gamma(h\to gg)$ with $\br_g^{\rm SM}\sim0.086$~\cite{HXSWGBR}:
\beq
\Gamma_h/\Gamma_h^{\rm SM}=1+\br_b^\sm(r_b-1) +\br_\tau^\sm(r_\tau-1)+ \br_g^\sm(r_t-1)\,,
\label{eq:Gammahtot}
\eeq
where we neglect the modification of the total width by the $\lambda_t$-induced change of $h\to\gamma\gamma$ due to the small $\br_\gamma^\sm\sim0.002$.
The total width is constrained as
$0.08\mev \leq \Gamma_h \leq 9.16\mev$ at $95\%$\,C.L.by CMS~\cite{Sirunyan:2019twz}, assuming an SM-like coupling structure, which applies in the considered framework. The SM prediction is
 $\Gamma_h^\sm =4.1\mev$.

\subsection{Single-flavor modification}
Consider the case that a single Yukawa coupling $\lambda_f$ is modified.
For $\mu_f \neq \hat \mu_f$ with
\begin{equation}
\hat \mu_f \equiv \frac{9}{1+8\br_f^\sm}\,,\label{eq:musingularity}
\end{equation}
the expression for the signal strength
\begin{align}\label{eq:muBR}
\mu_f = \frac{r_f}{1+\br_f^\sm(r_f-1)}
\end{align}
defines a circle in the $(T_R,T_I)$ plane
\begin{align}\label{eq:TImuBR}
{T_I^f}^2 + (T_R^f - T_{R0}^f)^2 = R_T^2 \,,
\end{align}
centered at $(T_R^f,T_I^f) = (T_{R0}^f,0)$, and with radius $R_T$ \footnote{In the vicinity of $\hat \mu_f$, the radius gets very large. Precisely at that value, the solutions of \eqref{eq:muBR} become independent of $T_I^f$, namely $T_R=-2/3$, or $\br_f^\sm=1$. However, such values are excluded by more than $3\sigma$ by the bounds of Eqs.~\eqref{eq:mutautaumeas} - \eqref{eq:mutexp}.}
\begin{align}\label{eq:radius}
\begin{split}
T_{R0}^f &= -\frac{3-\mu_f\,(1+2\br_f^\sm)}{9-\mu_f\,(1+8\br_f^\sm)} \,,\\
R_T^2 &= \frac{4\mu_f\,(1-\mu_f\,\br_f^\sm)(1-\br_f^\sm)}{\left[ 9-\mu_f\,(1+8\br_f^\sm) \right]^2}\,.
\end{split}
\end{align}
Thus, for a given $\br_f^\sm$, only a certain range of $\mu_f$ yields a real solution:
\begin{equation}
 \mu_f \leq \frac{1}{\br_f^\sm}\,. \label{eq:muboundary}
\end{equation}

It is interesting to note that, even for $\mu_{f}=1$, there exist solutions other than the trivial $T_R^f=T_I^f=0$ one, and they are independent of $\br_f^\sm$: $T_{R0}^f=-1/4$ and $R_T^f=1/4$ such that
\beq\label{eq:muone}
T_I^{f2}=-\frac12 T_R^f-T_R^{f2}.
\eeq
Thus, even if experiments close in on $\mu_{f}=1$, there will be an allowed circle in the $(T_R^f,T_I^f)$ plane. In particular, a new source of $CP$ violation, $T_I^f\neq0$
(with $|T_I^f|\leq1/4$), will be allowed. An experimental range, $\mu^{\rm min}\leq \mu \leq \mu^{\rm max}$, translates into an allowed region between two circles in this plane.

\subsubsection{$\lambda_\tau$}
The Yukawa coupling $\lambda_\tau$ is constrained by measurements of $\mu_{\tau^+\tau^-}$.
If only the $\tau$ Yukawa coupling is modified by dim-6 terms, then only $\Gamma(h\to \tau^+\tau^-)$ and $\Gamma_h$ are modified from their SM predictions. One can therefore combine the measurements of all Higgs production modes with the Higgs decaying into a pair of tau-leptons (for further details, see Table \ref{tab:muexp}):
\begin{equation}\label{eq:mutautaumeas}
 \mu_{\tau^+\tau^-}=0.91\pm 0.13.
\end{equation}
Using Eq.~(\ref{eq:muBR}) for $f=\tau$
where $r_\tau$ is defined in Eq.~(\ref{eq:defrf}), we find that both the upper and lower $2\sigma$-bounds yield a circle in the $(T_R^\tau, T_I^\tau)$ plane, resulting in the LHC-allowed ring shown in Fig.~\ref{fig:TRTI_tau_b}.

\subsubsection{$\lambda_b$}
The Yukawa coupling $\lambda_b$ is constrained by measurements of $\mu_{b\bar b}$ via Eq.~(\ref{eq:muBR}) for $f=b$.
Neglecting the 1\% bottom loop contribution to ggF, we combine all available production modes with the subsequent decay of $h\to b\bar b$ (see Table~\ref{tab:muexp}) as
\begin{equation}\label{eq:mubb}
 \mu_{b\bar b}=1.02\pm0.14\,,
\end{equation}
which is dominated by $\mu_{Vh}^{bb}$.

The fact that a modification of $\lambda_b$ affects not only $\Gamma(h\to b\bar b)$ but also the total width, $\Gamma_h$, has a significant impact on the resulting LHC-allowed ring in the $(T_R^b, T_I^b)$ plane, broadening it with respect to the case of a final state with a low branching ratio.

The constraint on $T_R^b, T_I^b$ from the total Higgs width is comparable to but weaker than from Eq.~(\ref{eq:mubb}) for negative $T_R^b$, and significantly weaker for positive $T_R^b$. Therefore, it is not shown in Fig.~\ref{fig:TRTI_tau_b}.

\subsubsection{$\lambda_t$}
The Yukawa coupling $\lambda_t$ is constrained by measurements of $\mu_{\rm ggF}$, $\mu_{tth}$ and $\mu_{\gamma\gamma}$. The latter provides a weaker constraint than the former two, so we do not use it. If the only modified Yukawa coupling is that of the top quark, then it is meaningful to combine the signal strengths of all of these top-mediated production processes, with all decays fixed to their SM values:
\begin{equation}\label{eq:mutexp}
\mu_{{\rm ggF}+t\bar th}=1.09 \pm 0.08\,,
\end{equation}
which is dominated by the ggF process. For details see Table~\ref{tab:muexpt}. Using Eq.~(\ref{eq:muBR}) for $f=t$ and $\br_f=\br_g$, the remarkable precision of the experimental range (\ref{eq:mutexp}) results in a narrow LHC-allowed ring in the $(T_R^t, T_I^t)$ plane, see Fig~\ref{fig:TRTI_t}.

\subsection{Two-flavor modification}
In the presence of two dim-6 Yukawa terms, the modifications of Higgs production and/or decay can be grouped into the two categories detailed in Table \ref{tab:categoriesf1f2}.
\begin{table}[b]
 \begin{center}
   \caption{Possible modifications of the signal strengths $\mu_I^F$ for two modified Yukawa couplings of the fermions $f_1$, $f_2$.
  The Higgs production cross section $ \sigma _I$, with $I=Vh$+VBF or ggF+$tth$, and/or the partial decay width $\Gamma (h\to F)$, with $F=f_i \bar f_i$ or $VV$, are modified by $ r_{f_i}$. SM denotes that the particular process is not modified. The total Higgs width $\Gamma_h$ is modified by both modified Yukawa couplings.}
   \begin{tabular}{|c|c|c||c|c||c|}
   \hline
   $\sigma_I$ &$\Gamma(h\to F)$ &$\Gamma_h$ &$f_1, f_2$ &process &dependence\\ \hline\hline
   \multirow{3}{*}{SM} &\multirow{3}{*}{$f_1$} &\multirow{3}{*}{$f_1, f_2$} &$\tau, b$ &any production, $h\to \tau\tau, b\bar b$& \multirow{4}{*}{A: Eq.~(\ref{eq:prodSM_decf1f2})} \\
   &&&$t,\tau$ &$Vh$+VBF, $h\to \tau\tau$&\\
   &&&$t,b$ &$Vh$+VBF, $h\to b \bar b$&\\ \cline{1-5}
   $f_1$&SM &$f_1, f_2$&$t, b/\tau$ &ggF+$tth$, $h\to VV$ &
   \\ \hline
\multirow{2}{*}{$f_1$} &\multirow{2}{*}{$f_2$} & \multirow{2}{*}{$f_1, f_2$} &$t,\tau$ &ggF+$tth$, $h\to \tau\tau$ & \multirow{2}{*}{B: Eq.~(\ref{eq:prodf1_decf1f2})} \\
&&&$t,b$ &ggF+$tth$, $h\to b\bar b$ &\\ \hline
  \end{tabular}
\label{tab:categoriesf1f2}
 \end{center}
\end{table}
The dependence of the signal strengths on the modified Yukawa interactions $r_f$ and on the SM branching ratios is given by
\begin{align}
  &\textrm{A:}~~&\mu_{\sm}^{f_1} =\mu^{\sm}_{f_1}
  &=\frac{r_{f_1}}{\Gamma_h/\Gamma_h^\sm}
  =\frac{r_{f_1}}{1+\br_{f_1}^\sm(r_{f_1}-1)+\br_{f_2}^\sm(r_{f_2}-1)} \label{eq:prodSM_decf1f2}\,,\\
&\textrm{B:}~~&\mu_{f_1}^{f_2}
&=\frac{r_{f_1}\,r_{f_2}}{\Gamma_h/\Gamma_h^\sm}
  = \frac{r_{f_1}\,r_{f_2}}{1+\br_{f_1}^\sm(r_{f_1}-1)+\br_{f_2}^\sm(r_{f_2}-1)}\label{eq:prodf1_decf1f2}\,,
\end{align}
with $f_1, f_2 = \tau, b, t$ and where $\br^\sm_t\equiv \br^\sm_g$.
Here the lower index of the signal strength $\mu$ denotes the modification of the Higgs production cross section whereas the upper index refers to the modification of the partial decay width, keeping in mind that the total Higgs width is modified by all modified Yukawa couplings.

In addition to the combinations A and B, we also evaluate the constraint from the total width itself with the dependence on $r_{f}$ given in Eq.~(\ref{eq:Gammahtot}). However, the present experimental bound on the total width~\cite{Sirunyan:2019twz} does not lead to to a constraint on $T_I^{f_1, f_2}$ exceeding the constraints from the signal strengths of cases A and B.

When constraining dim-6 Yukawa couplings of two fermions $f_1$, $f_2$ simultaneously, we are dealing with four SMEFT parameters. As an example of these constraints, in what follows we set $T_R^{f_1, f_2}=0$ and present the bounds in the $T_I^{f_1}-T_I^{f_2}$ plane. 
See Appendix \ref{app:TR=0} for details on how to obtain the corresponding limits.

\subsubsection{$\lambda_b$ and $\lambda_\tau$}
The production rates are neither affected by  $\lambda_b$ nor $\lambda_\tau$. Thus, we can still use the experimental ranges of Eq.~(\ref{eq:mutautaumeas}) for $\mu_{\tau^+\tau^-}$ and Eq.~(\ref{eq:mubb}) for
$\mu_{b\bar b}$. The theoretical expression for $\mu_{\tau^+\tau^-}$ and $\mu_{b\bar b}$ are, however, modified due to the modification of the total Higgs width by the two different Yukawa couplings,
see Eqs.~(\ref{eq:prodSM_decf1f2}), (\ref{eq:TI1a}) and (\ref{eq:TI2a}) with $f_1=\tau, b$ and $f_2=b,\tau$, respectively.
The total Higgs width constrains $|T_I^b|\lesssim 0.6$, i.e. similar to the direct $h\to b\bar b$ bound, but not stronger. Therefore, the $\Gamma_h$ bound is not shown in Fig.~\ref{fig:btau_TR0}.

\subsubsection{$\lambda_t$ and $\lambda_\tau$}
\label{sec:ttau}
For this combination of couplings, we use three relevant constraints:
\begin{align}
\begin{split}
\mu_{\rm ggF}^{\tau\tau}&=0.99 \pm 0.44\,,\\
\mu_{{\rm VBF}+Vh}^{\tau\tau}&= 1.09\pm 0.26\,,\\
\mu_{{\rm ggF}+tth}^{VV}&=1.08\pm0.08\,,
\end{split}
\end{align}
Because of the significantly higher precision of $\mu_{\rm ggF}^{\tau\tau}$ compared to $\mu_{tth}^{\tau\tau}$ (see Table ~\ref{tab:muexp}), we do not combine both production modes for the decay into $\tau\tau$, but use $\mu_{\rm ggF}^{\tau\tau}$. 
The theoretical expression for this channel
is given by Eq.~(\ref{eq:prodf1_decf1f2}) with $f_1=t$, $f_2=\tau$.

The combination of the $t$-independent production processes VBF and $Vh$, followed by the decay into $\tau^+\tau^-$, constrains $T_I^\tau$ with a mild dependence on $T_I^t$ due to the modification of the total Higgs width, see Eq.~(\ref{eq:prodSM_decf1f2}) with $f_1=\tau$.

The opposite combination of Higgs production via $\lambda_t$ and the decay into $VV$ mainly constrains $T_I^t$ with a mild dependence on $T_I^\tau$ via $\Gamma_h$. The structure of the signal strength $\mu_{t}^{VV}$ is given by Eq.~(\ref{eq:prodSM_decf1f2}) with $f_1=t$ and $f_2=\tau$.

\subsubsection{$\lambda_t$ and $\lambda_b$}
The combination of $T_I^t$ and $T_I^b$ is constrained by
\begin{align}\label{eq:tb}
\begin{split}
\mu_{t\bar th+ggF}^{b \bar b} &= 0.88 \pm 0.43\,,\\
\mu_{VH}^{bb}&=0.98\pm0.15\,,\\
\mu_{{\rm ggF}+tth}^{VV}&=1.08\pm0.08\,.
\end{split}
\end{align}
As $\mu_{\rm VBF}^{bb}$ is not available at comparable precision, there is no need for a combination of VBF and $Vh$ to constrain the $b$ Yukawa coupling.
The corresponding theoretical expressions are
$\mu_{{\rm ggF}+tth}^{b \bar b}$ according to Eq.~(\ref{eq:prodf1_decf1f2}) with $f_1=t$, $f_2=b$,
$\mu_{Vh}^{b \bar b}$ according to Eq.~(\ref{eq:prodSM_decf1f2}) with $f_1=b$, $f_2=t$, and
$\mu_{{\rm ggF}+tth}^{VV}$ according to Eq.~(\ref{eq:prodSM_decf1f2}) with $f_1=t$, $f_2=b$. 
The limits on $(T_I^\tau, T_I^b)$ resulting from $\mu_{Vh}^{bb}$ are weaker than from the other two experimental processes and are therefore not shown in Fig.~\ref{fig:tb_TR0}.

\section{Results}

\subsection{Single flavor modification}
\label{sec:results1}
We present in this section the results of our combined analysis of three physical observables: $Y_B$, $d_e$ and $\mu_{f}$,  from a single flavor source. We note the following points:
\begin{itemize}
\item Both the baryon asymmetry and the electron EDM are proportional to $(y_f/y_f^{SM})^2 T_I^f$, except for the top quark. This implies that, for a single $CP$ violating source from $f\neq t$, contours of constant $Y_B$ are also contours of constant $d_e$. In contrast, $Y_B^t$ is approximately constant in $T_R^t$ due to the large Yukawa coupling contributing to its thermal mass.
\item The $Y_B$ dependence on $T_R^f$ is mild. Negative values of $T_R$ generate a larger baryon asymmetry.
\item The value $\mu_{f}=1$ defines a circle in this plane through the SM point $T_I^f=T_R^f=0$.
\item Experimental bounds on $\mu_{f}$ constrain the dim-6 operators of each species to an annulus in the $ T_R^f,T_I^f $ plane.
\item As all signal strengths $\mu_f$ are compatible with 1, the radius is approximately $0.25$, where the exact value and the width of the ring depend on the precise bounds on $\mu_f$ and the value of $\br_f^\sm$, see Eq.~(\ref{eq:radius}). Hence the collider sensitivity on $T_R^f$ and $T_I^f$ reaches few times $\mathcal{O}(0.1)$.
\end{itemize}

\subsubsection{$\lambda_\tau$}
\begin{figure}[t]
 \begin{center}
\includegraphics[width=0.48\textwidth]{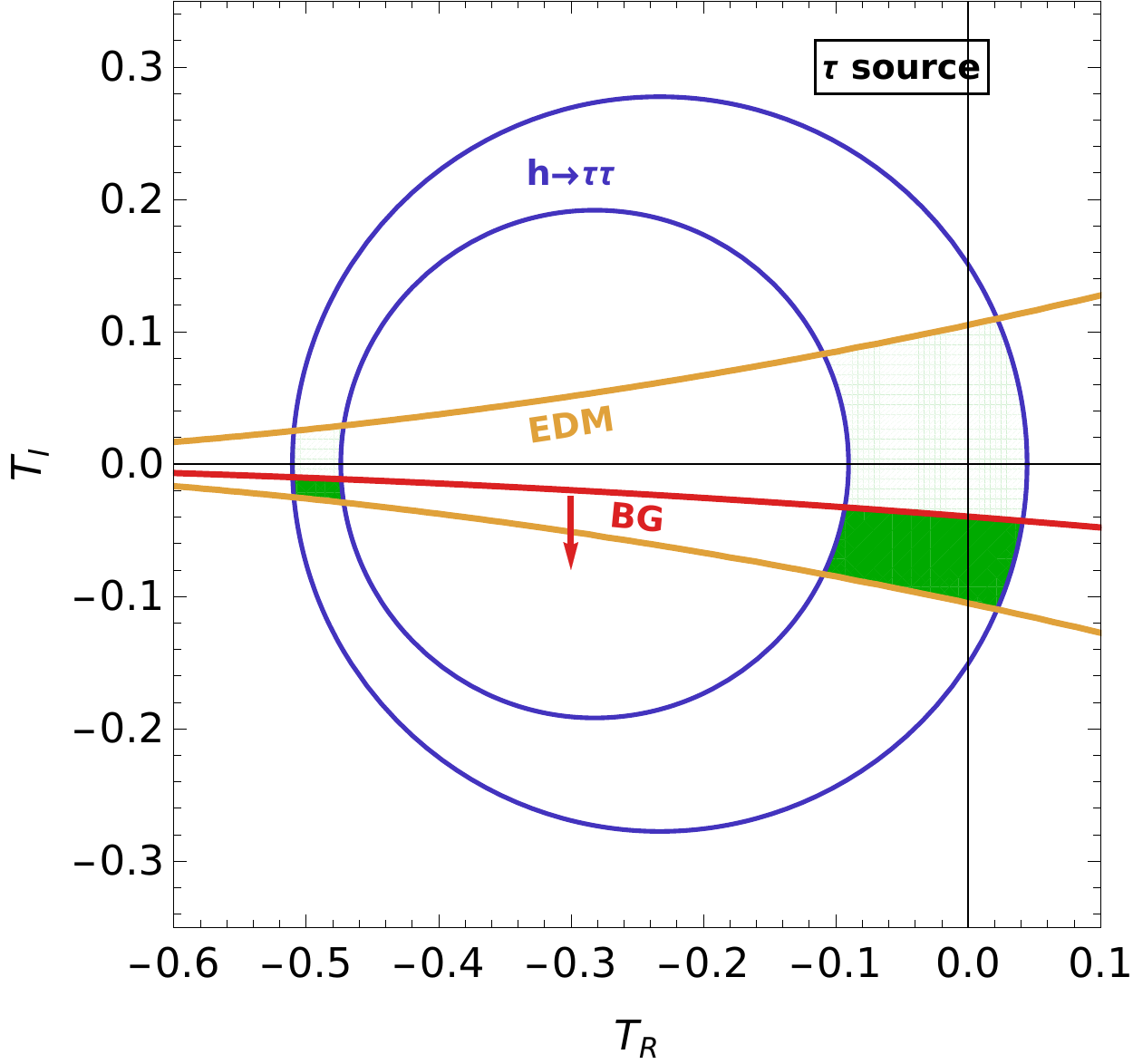}\hfill
\includegraphics[width=0.48\textwidth]{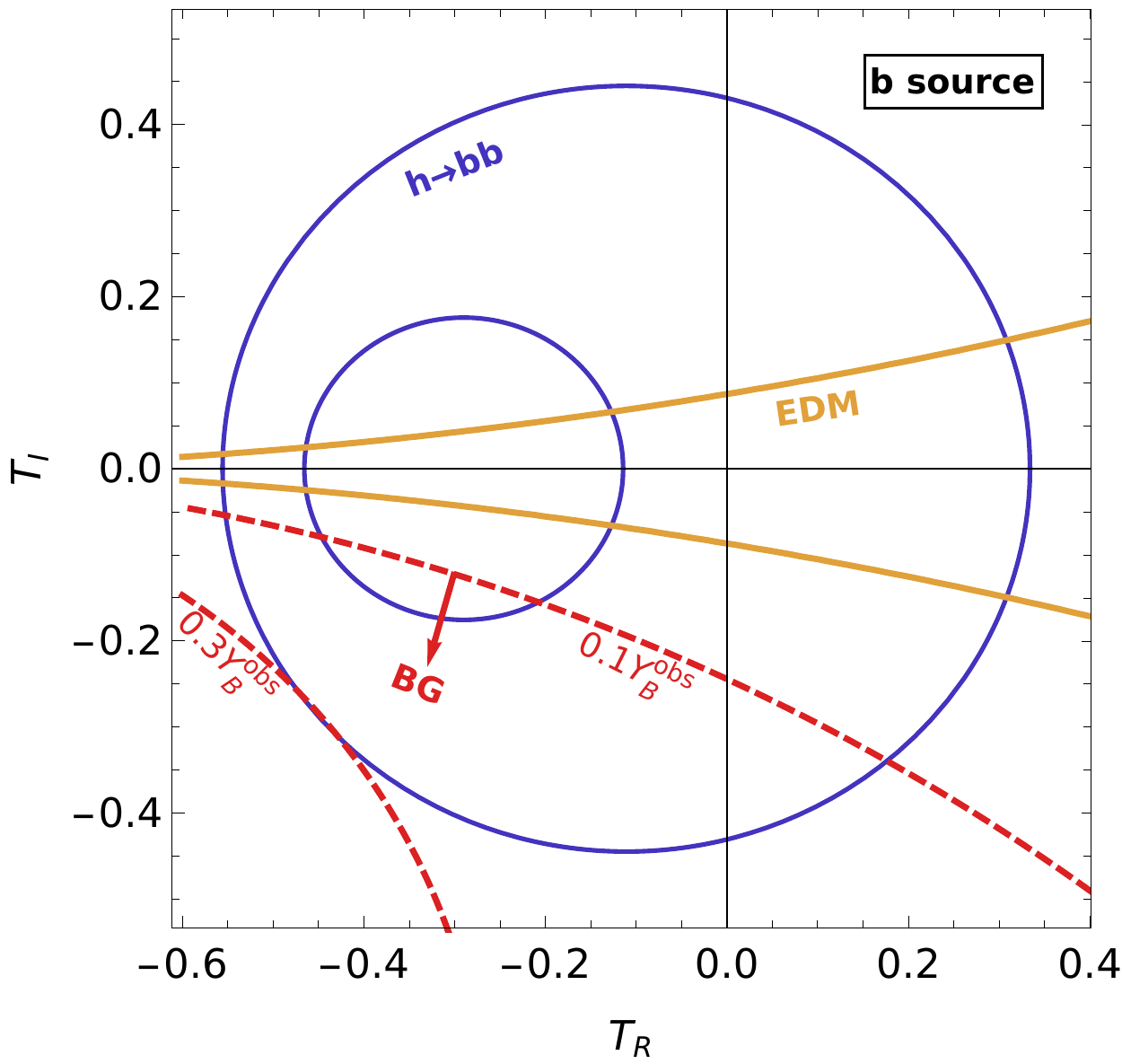}
  \caption{Constraints on $(T_R^{f}, T_I^{f})$, $f=\tau, b$, from $\mu_{f\bar f}$ (blue), $d_e$ (yellow) and $Y_B$ (red). Solid lines represent bounds, dashed red lines represent iso-$Y_B$ curves within the bound.
  Regions allowed by all constraints are highlighted in green.
  \textit{Left:} $\tau$-lepton source,
  \textit{right:} $b$-quark source.
 }
  \label{fig:TRTI_tau_b}
 \end{center}
\end{figure}

The constraints on $(T_R^\tau,T_I^\tau)$ are presented in Fig.~\ref{fig:TRTI_tau_b} (left). The constraints on a complex Yukawa coupling for the tau-lepton from $\mu_{\tau^+\tau^-}$ and from $d_e$ are comparable. While the EDM is more constraining on $T_I^\tau$, the decay rate for $ h\to \tau ^+\tau ^- $ restricts $T_R^\tau$.
Within the region allowed by the two measurements, there is a region where a complex $\lambda_\tau$ can generate enough $CP$ violation to account for the observed BAU. The largest value of $Y_B$ obtained in the allowed regions is
\begin{equation}
 Y_B^{\tau,{\rm max}}\simeq 2.4Y_B^{\rm obs}\,.\label{eq:YBtaumax}
\end{equation}

We quote this upper bound which is larger than $Y_B^{\rm obs}$ (and similar bounds further below) for three reasons:
\begin{itemize}
\item In our calculations, we use bubble wall parameters that are optimal for generating $Y_B$. The upper bound implies by how much these parameters can be less than optimal, and yet a complex $\lambda_\tau$ can provide the $CP$-violation necessary for baryogenesis.
\item Similarly, the upper bound gives a sense for how sensitive our conclusions are with respect to uncertainties and approximations in the $Y_B$ calculation.
\item The upper bound is informative on which future experiments can test this scenario in a definitive way.
\end{itemize}

\subsubsection{$\lambda_b$}
The constraints on $(T_R^b,T_I^b)$ are presented in Fig.~\ref{fig:TRTI_tau_b} (right).
Generating sufficient $CP$ violation from a complex $\lambda_b$ requires $\abs{T_I^b}>1$ (see Eq.~(\ref{eq:YBttaub})). Therefore we conclude that $\lambda_b$ cannot serve as the only source of $CP$ violation to account for $Y_B^{\rm obs}$. While the $\mu_{b\bar b}$ constraint alone allows $Y_B^{(b)}\leq0.33Y_B^{\rm obs}$, the $d_e$ constraint is stricter on $T_I^b$, leading to
\beq
Y_B^{(b)}\leq0.04Y_B^{\rm obs}.
\eeq

\subsubsection{$\lambda_t$}
\begin{figure}[ht]
 \begin{center}
\includegraphics[height=0.46\textwidth]{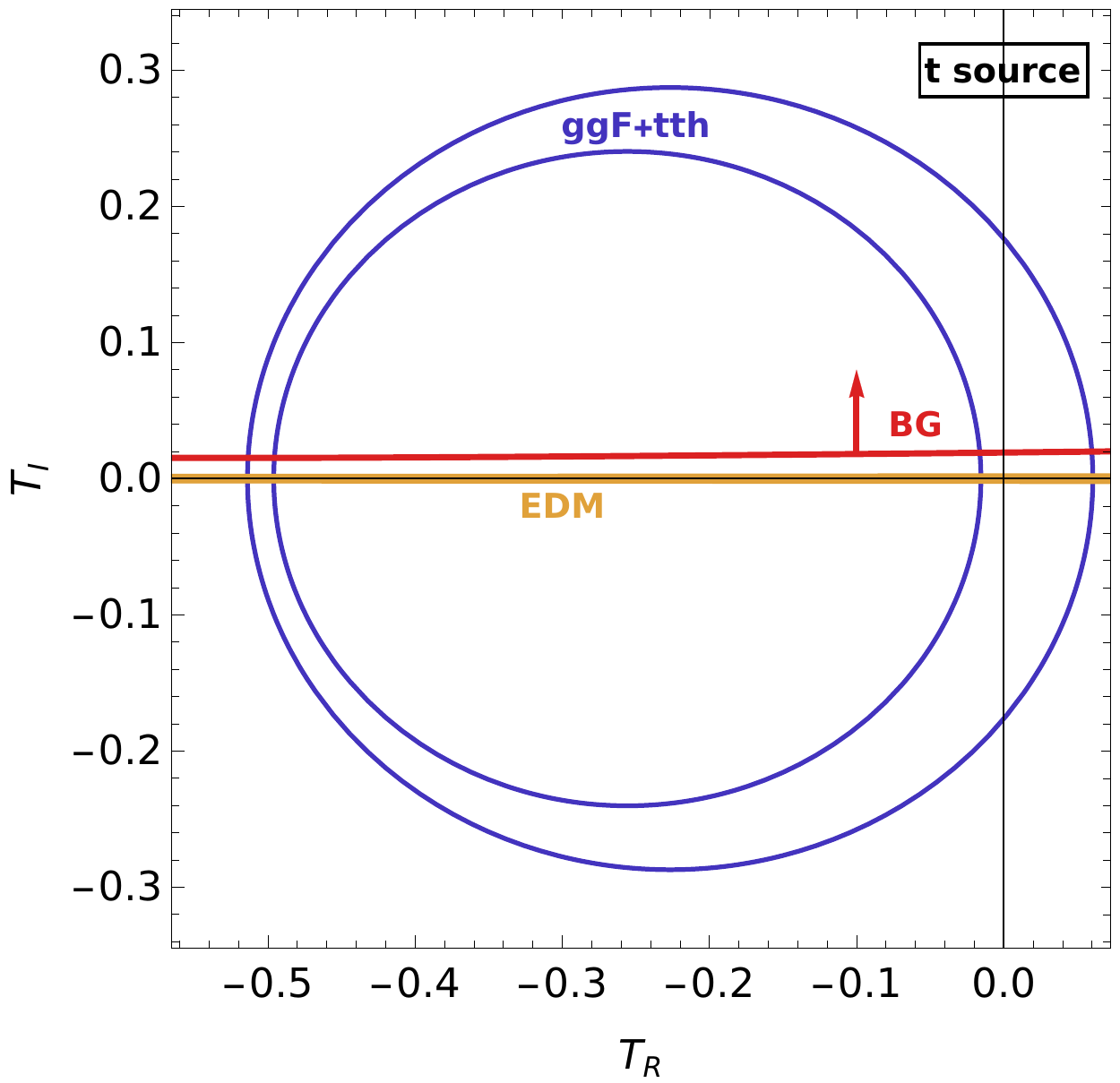}\hfill
\includegraphics[height=0.46\textwidth]{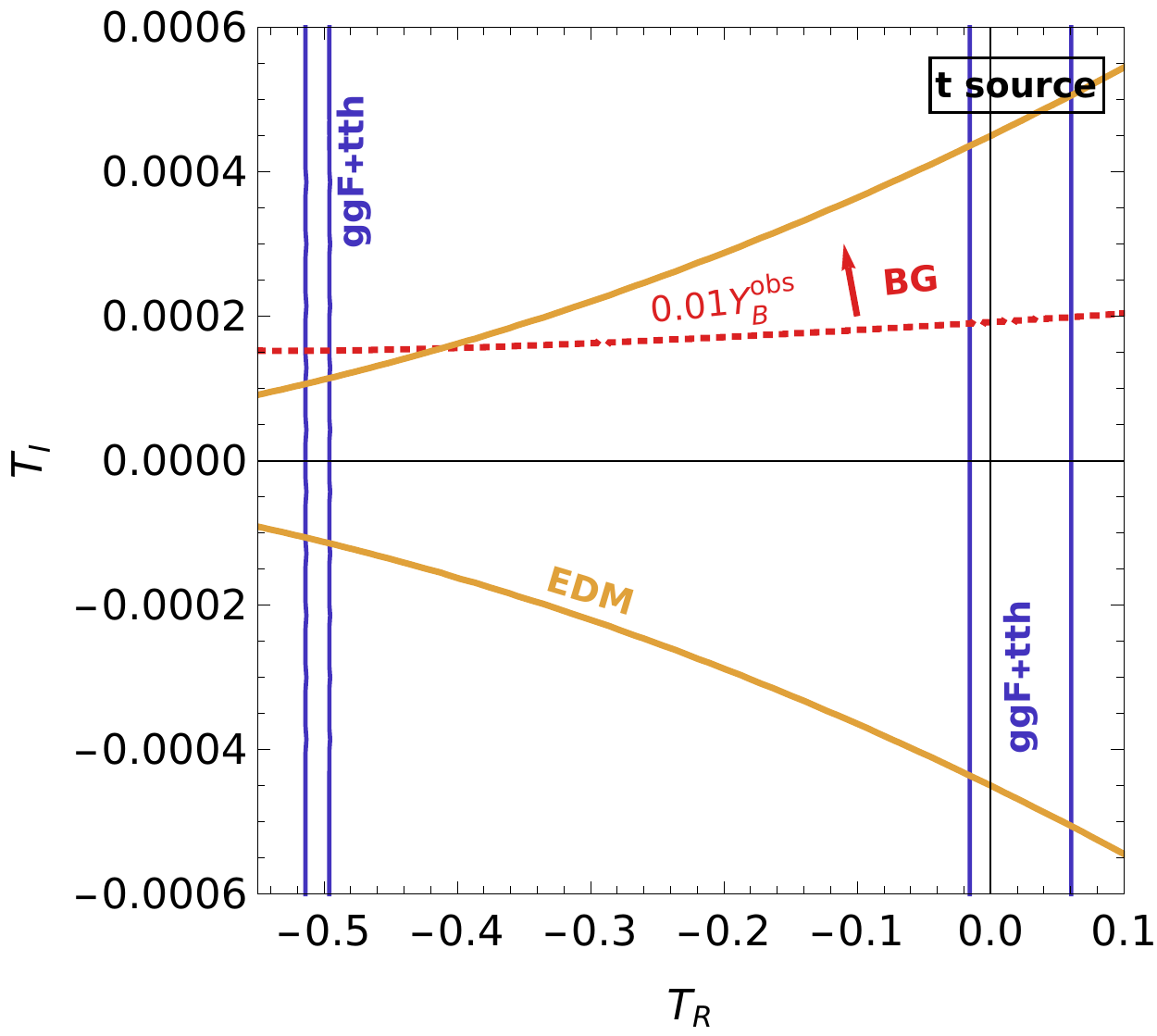}
  \caption{Top source: Constraints on $(T_R^{t}, T_I^{t})$
  from $\mu_{ggF+t\bar th+th}=1.09\pm0.08$ (blue), $d_e$ (yellow) and $Y_B$ (red).
  \textit{Left:} full range of LHC bound,
  \textit{right:} zoomed into range near EDM bound. The difference in the scaling of $ Y_B $ and $ d_e $ in this case is due to the non-negligible contribution of the Yukawa interactions to the thermal mass of the top.
  }
  \label{fig:TRTI_t}
 \end{center}
\end{figure}
The constraints on $(T_R^t,T_I^t)$ are presented in Fig.~\ref{fig:TRTI_t}.
The constraint on $T_I^t$ from $\mu_{{\rm ggF}+tth}$ is three orders of magnitude weaker than from $d_e$.
Thus, while the former by itself would allow for $Y_B^{(t)}\sim Y_B^{\rm obs}$, the latter restricts it:
\beq
Y_B^{(t)}\leq0.02Y_B^{\rm obs}.
\eeq
The right plot presents a zoomed-in version of the figure to illustrate the strength of the EDM constraint, reducing the collider ring to two strips, one around the origin, and a narrow one around $T_R^t\simeq -0.5$.
The latter translates to $\Lambda/\sqrt{X_R^t}\simeq v$, thus the EFT assumptions might not be fulfilled in this particular region of large negative $T_R^t$ whereas the cutoff scale is well above $v$ in the strip around the origin.
The calculation of $Y_B^{(t)}$ is affected by larger uncertainties than in the $\tau$- and $b$-cases due to the less suppressed higher-dimensional terms and the non-negligible higher-order effects in the VEV-insertion approximation~\cite{Postma:2019scv}. Nevertheless, as our prediction of $Y_B^{(t)}$ is $\mathcal{O}(100)$ away from $Y_B^{\rm obs}$, our qualitative observation remains that the top quark does not induce sufficient baryon asymmetry.

\subsection{Two flavor modification}
\label{sec:results2}
In general, we expect that the dimension-six terms in the SMEFT would modify all Yukawa couplings. A combined contribution to the various observables from two or more flavors may open up new regions in the parameter space that can account for the baryon asymmetry. To understand this statement, consider Eqs.~(\ref{eq: d_e coefficients}) and (\ref{eq:YBttaub}). Given the fact that $d_e$ and $Y_B$ depend on different linear combinations of $T_I^t$, $T_I^b$ and $T_I^\tau$, there could be simultaneously cancellations between their contributions to $d_e$ and enhancements of their contributions to $Y_B$.
Our choice of $T_R^f=0$ in the combination plots of two flavors represents a conservative case for the yield of $Y_B$.

\subsubsection{$\lambda_b$ and $\lambda_\tau$}
The constraints in the $(T_I^\tau,T_I^b)$ plane, for $T_R^\tau=T_R^b=0$, are presented in Fig.~\ref{fig:btau_TR0}. A much larger range of negative $T_I^\tau$ is consistent with $Y_B^{(\tau)}\gsim Y_B^{\rm obs}$. Here, $T_I^b$ cancels against $T_I^\tau$ in $d_e$, and maintains $\mu_{\tau^+\tau^-}$ within bounds by increasing the total width, $\Gamma_h$.
The maximal baryon asymmetry reached within the allowed range is
\begin{equation}
 Y_B^{b+\tau,{\rm max}}(T_I^\tau=-0.4, T_I^b=+0.4)\simeq 7.8 Y_B^{\rm obs}\,.
\end{equation}

\subsubsection{$\lambda_t$ and $\lambda_\tau$}
\begin{figure}
\centering
\includegraphics[width=0.48\linewidth]{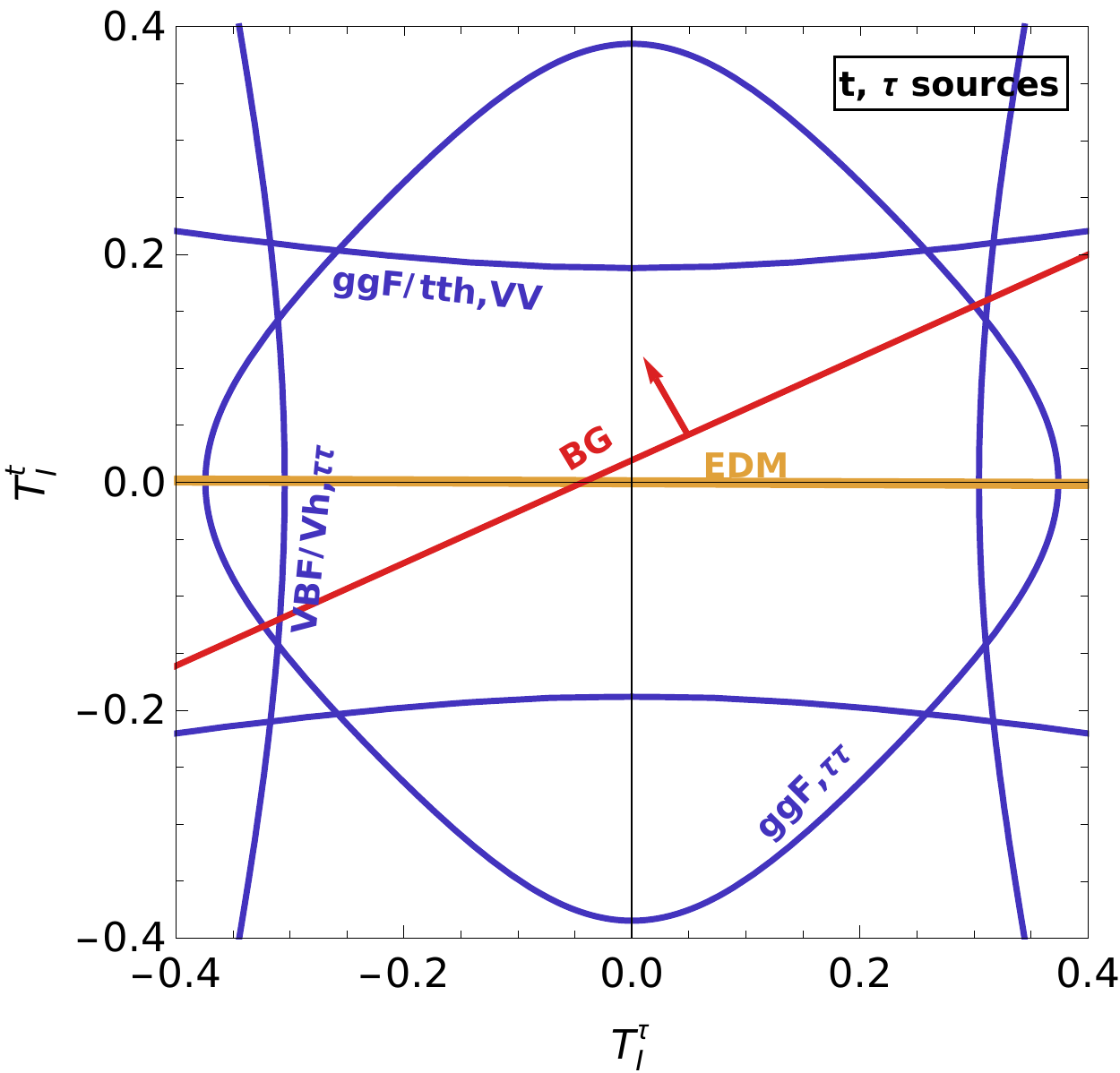} \hfill
\includegraphics[width=0.48\linewidth]{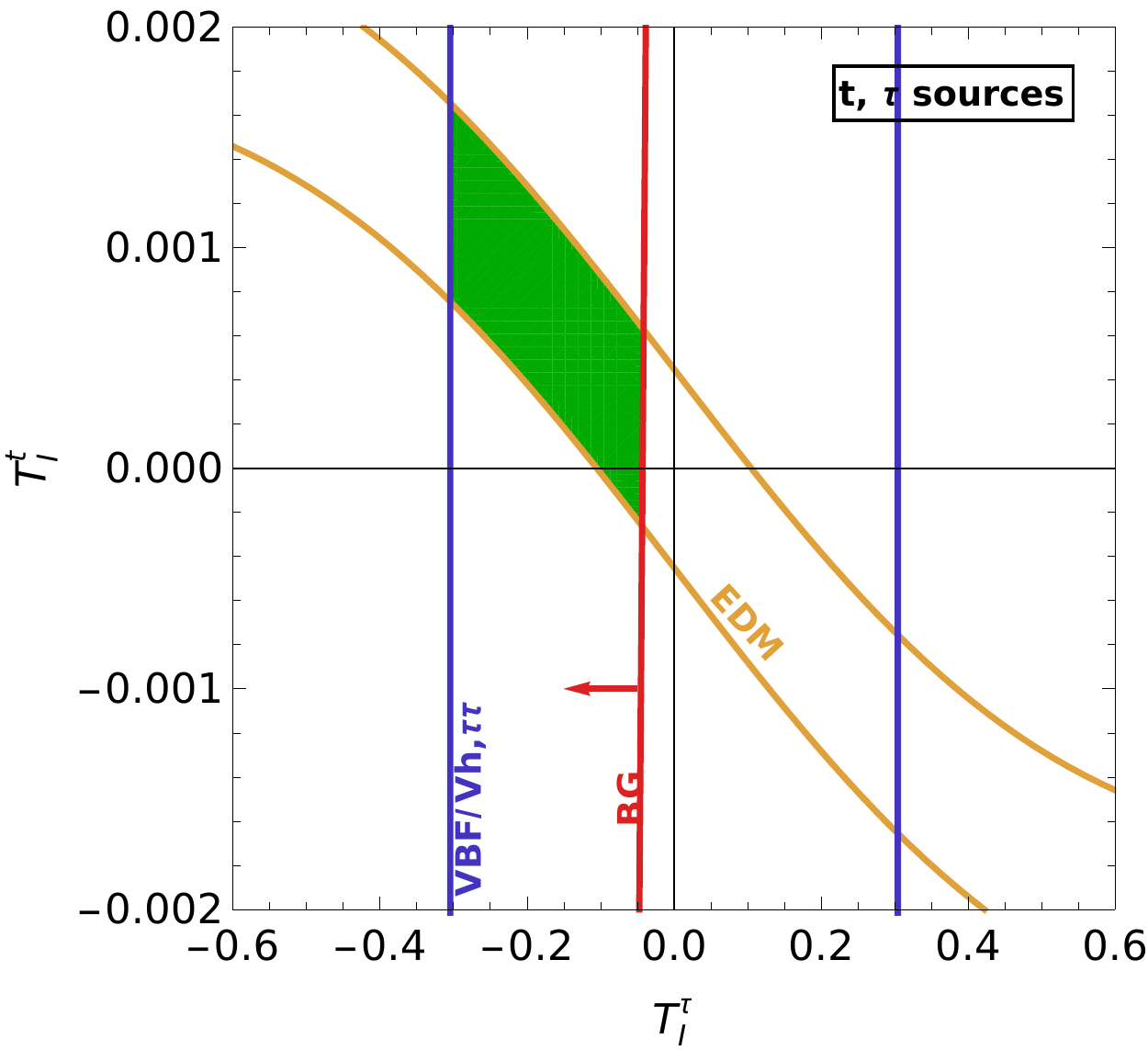}
\caption{Constraints on $t$ and $\tau$ sources with $T_R^{t,\tau}=0$ from the LHC (blue), the eEDM (yellow) and $Y_B$ (red). The parameter space allowed by all three constraints is highlighted in green. \textit{Left:} Full collider range allowed by $\mu_{{\rm ggF}}^{\tau\tau}=0.99\pm0.44$
and $\mu_{{\rm VBF}+Vh}^{\tau\tau}=1.09\pm0.26$
(combining ATLAS and CMS), \textit{right:} zoomed into the EDM-allowed region.
}
\label{fig:ttau_TR0}
\end{figure}

The constraints in the $(T_I^\tau,T_I^t)$ plane, for $T_R^\tau=T_R^t=0$, are presented in Fig.~\ref{fig:ttau_TR0}. A much larger range of negative $T_I^\tau$ is consistent with $Y_B^{(\tau)}\gsim Y_B^{\rm obs}$. Here, $T_I^t$ cancels against $T_I^\tau$ in $d_e$, and maintains $\mu_{\tau^+\tau^-}$ within bounds by decreasing the ggF and $tth$ production rates.
The maximal baryon asymmetry reached within the allowed range is
\begin{equation}
 Y_B^{t+\tau,{\rm max}}= Y_B^{t+\tau}(T_I^\tau=-0.3, T_I^t=+0.0016)\simeq 6.4Y_B^{\rm obs}\,.
\end{equation}

\subsubsection{$\lambda_t$ and $\lambda_b$}
\begin{figure}
\centering
\includegraphics[width=0.48\linewidth]{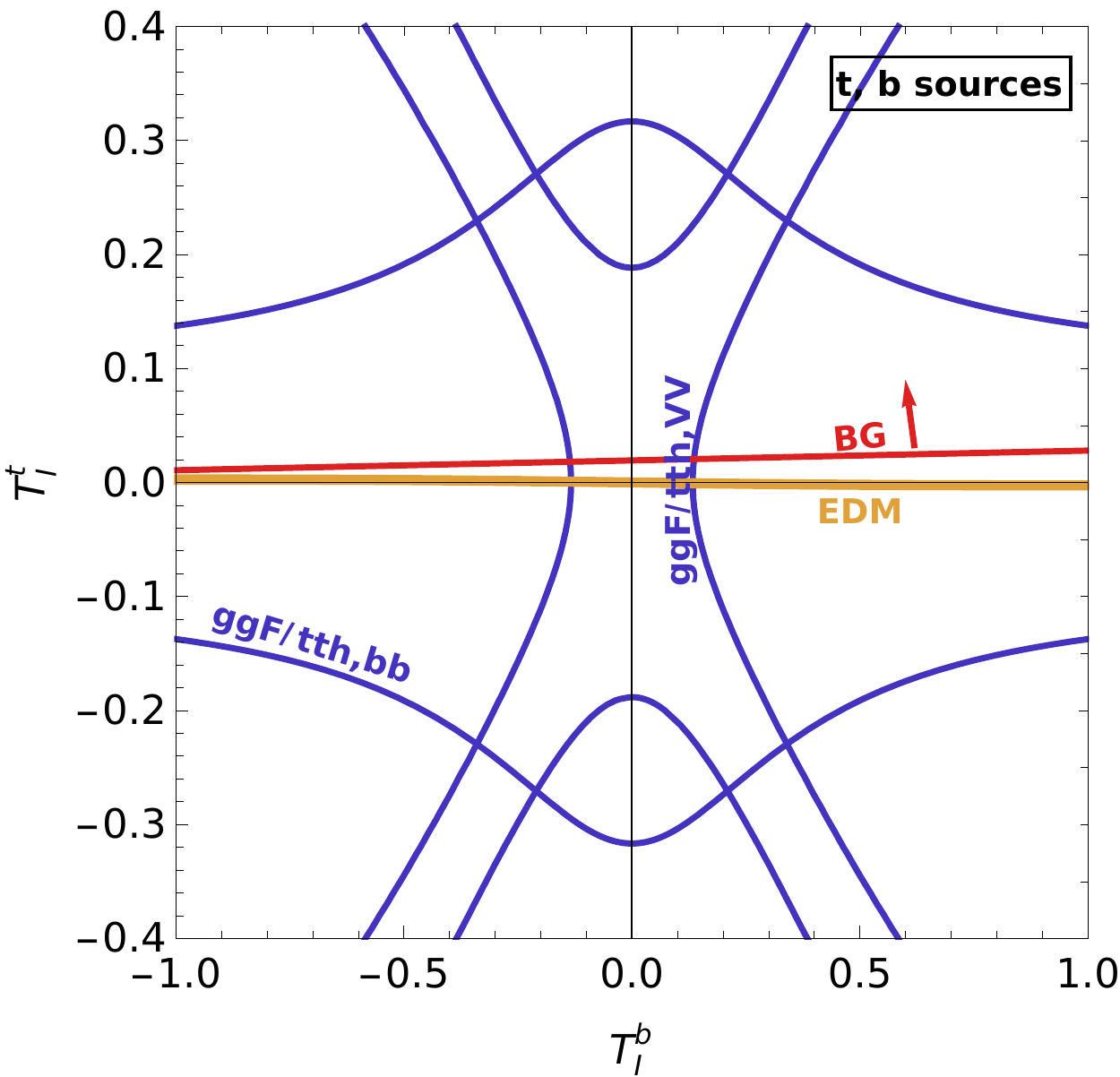} \hfill
\includegraphics[width=0.48\linewidth]{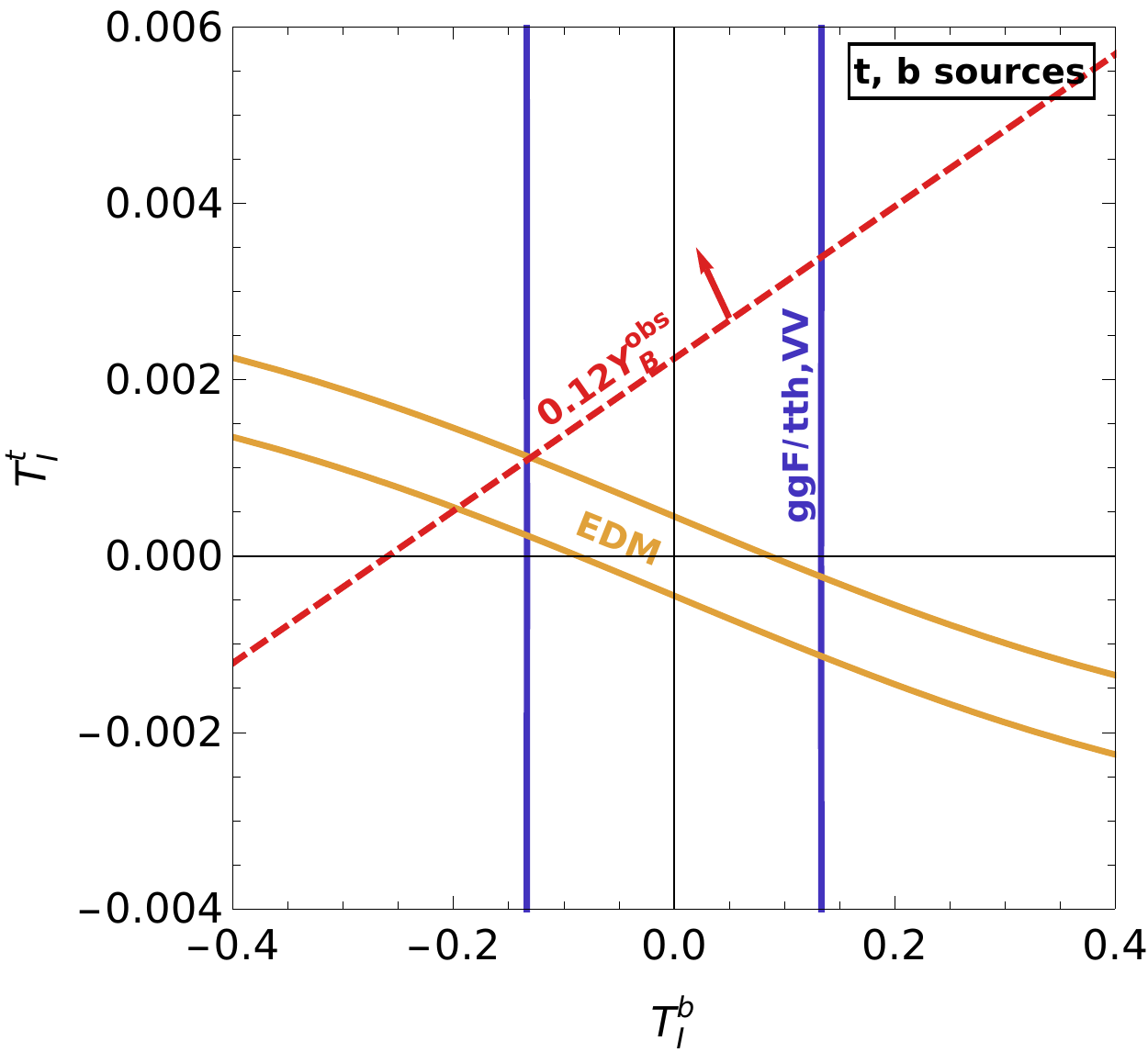}
\caption{Constraints on $t$ and $b$ sources with $T_R^{t,b}=0$ from the LHC (blue), the eEDM (yellow) and $Y_B$ (red). \textit{Left:} Full collider range (combining ATLAS and CMS) allowed by $\mu_{tth+ggF}^{bb}=0.88\pm0.43$ and $\mu_{ggF+tth+th}^{VV}=1.08\pm0.08$, \textit{right:} zoomed into the EDM-allowed region.
}
\label{fig:tb_TR0}
\end{figure}

The constraints in the $(T_I^b,T_I^t)$ plane, for $T_R^b=T_R^t=0$, are presented in Fig.~\ref{fig:tb_TR0}. The LHC bounds on $\mu_{{\rm ggF}+tth}^{VV}$ in combination with the electron EDM constrain $T_I^b$ to within a range such that
\beq
Y_B^{(t+b)}\lsim 0.12Y_B^{\rm obs}.
\eeq

\begin{figure}
\centering
\includegraphics[width=0.5\linewidth]{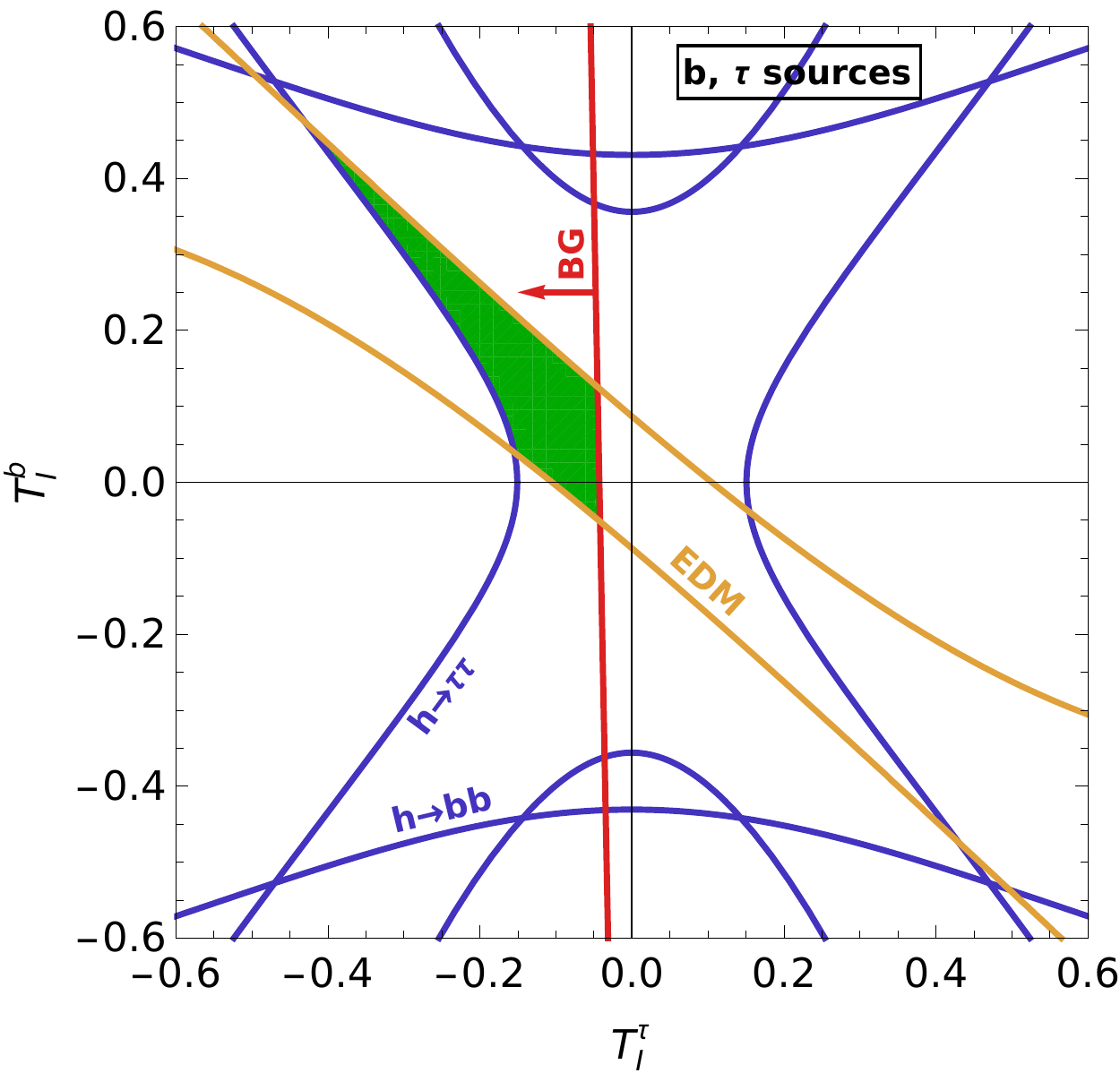} 
\caption{Constraints on $b$ and $\tau$ sources with $T_R^{b,\tau}=0$ from the LHC (blue), the eEDM (yellow) and $Y_B$ (red). The parameter space allowed by all three constraints is highlighted in green. Collider range allowed by $\mu_{\tau^+\tau^-}=0.91\pm 0.13$ and  $\mu_{b\bar b}=1.02\pm0.14$, with $\mu_I^F$ given by equation (\ref{eq:prodSM_decf1f2})
}
\label{fig:btau_TR0}
\end{figure}

\subsection{Obtaining $Y_B=Y_B^{\rm obs}$ with $d_e\simeq0$ and $\mu_I^F\simeq1$}
Even if experiments strengthen the upper bound on $d_e$ significantly, and narrow the allowed ranges around $\mu_I^F=1$ in all modes, the possibility that our SMEFT framework accounts for the $CP$ violation that is necessary for $Y_B=Y_B^{\rm obs}$ will remain viable.

To explain this statement, we go beyond the $T_R^f=0$ examples of the previous subsections. Consider, for example, the case that both $\lambda_\tau$ and $\lambda_b$ are modified by the dim-6 terms. We impose three constraints -- $d_e=0$, $\mu_{b\bar b}=1$ and $\mu_{\tau^+\tau^-}=1$ -- on the four parameters $T_{I,R}^{b,\tau}$. It is always possible to choose a combination of $T_I^\tau$ and $T_I^b$ such that $d_e=0$, and corresponding values of $T_R^\tau$ and $T_R^b$ such that $\mu_b=\mu_\tau=1$. We therefore have one free parameter, bounded within some range of values, and we can check whether in this range we can produce the observed baryon asymmetry. Using \eqref{eq: d_e coefficients} and \eqref{eq:prodSM_decf1f2}, we find that we can obtain up to
\begin{align}
Y_B^{b+\tau,{\rm max}}(d_e=0,\mu_b=\mu_\tau=1) = 10.25 Y_B^{\text{obs}} \,.
\end{align}
This corresponds to being very close to the point along the $ \mu _\tau =1 $ circle that maximizes the baryon asymmetry, while the $ b $ only slightly reduces the produced baryon asymmetry. Indeed, simply maximizing the baryon asymmetry along $ \mu _\tau =1 $ with $ T_I^b = T_R^b = 0 $ gives
$Y_B^{(b+\tau),{\rm max}} = 10.33 Y_B^{\text{obs}}$.

\section{Conclusions}
\label{sec:conclusion}
We studied in detail aspects of $CP$ violation in the SMEFT. Specifically, we considered dimension-six terms involving the Higgs field and the third generation fermion fields, and analyzed the consequences for electroweak baryogenesis, for the electric dipole moment of the electron, and for Higgs production and decay rates. There is an interesting interplay between the three constraints. Our main conclusions are the following:
\begin{itemize}
\item A complex Yukawa coupling of the tau-lepton can provide large enough $CP$ violation to account for the baryon asymmetry.
\item A complex Yukawa coupling of neither the top-quark quark nor the bottom-quark can provide large enough $CP$ violation to account for the baryon asymmetry: $Y_B^{(t)}\lsim0.02 Y_B^{\rm obs}$ and $Y_B^{(b)}\lsim0.04 Y_B^{\rm obs}$.
\item The reason why the tau-lepton is more successful than the bottom- or the top-quark is that the strong sphalerons do not act on the tau-lepton. Therefore, the asymmetry created by a complex tau-Yukawa coupling is not washed out as strongly as the one of the quarks. Furthermore, the large leptonic diffusion coefficients, see Eq.~\ref{eq:diffusionC}, enable an efficient diffusion of the asymmetry into the broken phase. These effects overcompensate the smaller, but still sizeable, $\tau$-Yukawa coupling.
\item Limited by the upper bound on $\mu_{\mu^+\mu^-}$, a complex Yukawa coupling of the muon can account for $0.16 Y_B^{\rm obs}$~\cite{Fuchs:2019ore}, i.e. for more than the third-generations quarks, due to the advantages of leptons as for the $\tau$ despite the smaller $y_\mu^\sm$. Even a future measurement of $\mu_{\mu^+\mu^-}=1$ would allow for $0.12 Y_B^{\rm obs}$, which remains more than from $t$ or $b$.
\item There can be substantial cancellations between the contributions of the third generation fermions to $d_e$, that are not necessarily accompanied by cancellations in their contributions to $Y_B$. In fact, one can have:
     \begin{itemize}
     \item $d_e^{(b+\tau)}=0$ simultaneously with $Y_B^{(b+\tau)}\sim Y_B^{\rm obs}$;
     \item $d_e^{(t+\tau)}=0$ simultaneously with $Y_B^{(t+\tau)}\sim Y_B^{\rm obs}$;
     \item $d_e^{(t+b)}=0$ simultaneously with $Y_B^{(t+b)}\sim 0.12Y_B^{\rm obs}$.
     \end{itemize}
     In other words, we can have successful electroweak baryogenesis without having a signal for the electron dipole moment of the electron.
\item Such cancellations allow $T_I^\tau$ to further increase $Y_B^{(\tau)}>Y_B^{\rm obs}$. Consequently, near-future measurements of $d_e$ or $\mu_{\tau^+\tau^-}$ are unlikely to exclude the scenario of $\tau$-driven electroweak baryogenesis.
\item In fact, even if future experimental measurements establish $d_e\simeq0$ and $\mu_I^F\simeq1$ in all relevant modes, the scenario where a complex $\lambda_\tau$ provides the $CP$ violation that is necessary for baryogenesis will not be excluded.
\item Measuring $CP$ violation in the decays of $h\to\tau^+\tau^-$ is crucial to determine the viability of baryogenesis in this SMEFT framework with dimension six couplings.
\end{itemize}

For a complex $\lambda_\tau$ to account for $Y_B^{\rm obs}$, $T_I^\tau$ in the range $0.01-0.1$ is required. This can be translated via $T_I^\tau\equiv(v^2/2\Lambda^2)(X_I^\tau/y_\tau)$ into a upper bound,
\beq
\Lambda/\sqrt{X_I^\tau}\lsim18\ {\rm TeV}\ (0.01/T_I^\tau)^{1/2}.
\eeq

Our work is largely consistent with Refs.~\cite{deVries:2017ncy,deVries:2018tgs}.
We extended  previous results to consider non-negligible values of $T_R^f$ which enlarges the available parameter space and provides a more complex interplay among the different fermion species when applying the constraints from EDMs and Higgs physics results from colliders. Moreover, we consider also the cases that two third generation fermions contribute to the $ CP $ violation, and find that even if experiments find no deviations from the SM predictions, large enough $Y_B$ can be generated in the SMEFT.

It is interesting to note that, while $d_e$ and $Y_B$ sum over the contributions of all Yukawa couplings, the ATLAS and CMS measurements of $\mu_I^F$ are flavor specific. Moreover, a dedicated search for $CP$ violation in the decay $h\to f\bar f$ is unique in allowing a separate investigation of the imaginary part of each Yukawa coupling $\lambda_f$.

We have shown that the $\tau$ interactions are the only stand-alone  option for generating the matter-antimatter asymmetry of the Universe. This situation implies that a significant step in probing the $CP$ violation aspect of electroweak baryogenesis in the SMEFT framework can be achieved by searching for $CP$ violation in $h\to\tau^+\tau^-$ decays. Thus, $CP$ violation properties of the $\tau$-lepton interactions should be a priority experimentally as this could provide the most significant constraints on the viability of electroweak baryogenesis in the context of the SMEFT framework.

In the future it will be important to further improve the sensitivity to a possible $CP$-odd component of this coupling at the LHC and future colliders, and to evaluate the baryon asymmetry corresponding to the constrained amount of $CP$-violation. 
An experimental HL-LHC projection~\cite{ATL-PHYS-PUB-2019-008}
and several phenomenological analyses~\cite{Harnik:2013aja,Berge:2014sra,Hagiwara:2016zqz,Chen:2017nxp} have already been performed in this direction. Machine Learning (ML) may also play a useful role in further scrutinizing the $CP$ nature of the Higgs-$\tau$ interaction~\cite{Richter-Was:2019xiy,Lasocha:2020ctd}. In addition, $CP$ analyses of the $t$-Higgs coupling have been proposed, including ML~\cite{Ren:2019xhp,Englert:2019xhk,Faroughy:2019ird}.

\acknowledgments{We are grateful to Jorinde van de Vis for very helpful discussions. We thank Daniel Aloni for interesting discussions during the early stage of this work and the Fermilab and Cornell theory seminar audiences for stimulating comments. ML would like to deeply thank the Weizmann Institute of Science for its hospitality during the completion of this work.
EF was supported by the Minerva Foundation.
YN is the Amos de-Shalit chair of theoretical physics, and is supported by grants from the Israel Science Foundation (grant number 394/16), the United States-Israel Binational Science Foundation (BSF), Jerusalem, Israel (grant number 2014230), and the Yeda-Sela (YeS) Center for Basic Research.
}

\appendix
\section{Collider limits}
\subsection{Experimental bounds}
\label{app:munumbers}
Tables \ref{tab:muexp} and \ref{tab:top} contain the details of the collider limits from Higgs physics, by both CMS and ATLAS, that have been used to constrain the Higgs signal strengths for all third generation fermions.
\renewcommand{\arraystretch}{0.8}
\begin{table}[h]
 \begin{center}
 \caption{Collider limits on the Higgs signal strengths involving the $\tau$, $\mu$ and $b$.
 In our private combinations, the uncertainties are approximated as symmetric.
}
  \begin{tabular}{||c||c|c|c|c|c||c||}
   \hline \hline
   channel &experiment	&$\sqrt{s}/\tev$ &$\mathscr{L}/\ifb$ &comment 	&$\mu$	&Ref\\ \hline \hline
   \multirow{4}{*}{$h\to \tau^+\tau^-$}	&ATLAS+CMS	&7+8	&$5+20$	&&$1.11^{+0.24}_{-0.22}$	 &\cite{Khachatryan:2016vau}\\
  	&ATLAS	&13	&36.1	&ggF, VBF	&$1.09^{+0.35}_{-0.30}$&\cite{Aaboud:2018pen}\\
 	&CMS	&13	&77	&ggF, $\bar bb$, VBF, $Vh$	&$0.75\pm0.17$&\cite{CMS:2019pyn}\\ \cline{2-7}
  	&ATLAS+CMS	&7+8+13	&		&all prod., \pc	&$0.91\pm0.13$	 &\cite{Khachatryan:2016vau,Aaboud:2018pen,CMS:2019pyn}\\	
 	\hline
    \multirow{2}{*}{$h\to \mu^+\mu^-$}	&ATLAS	&\multirow{2}{*}{13}	&139	&\multirow{2}{*}{upper bound at 95\% C.\,L.}	&$<1.7$&\cite{ATLAS:2019ain}\\
 	&CMS	&	&35.9	&	&$<2.9$&\cite{Sirunyan:2018hbu}\\\hline	
    \multirow{6}{*}{$h\to \bar bb$}	&\multirow{2}{*}{ATLAS}	&\multirow{2}{*}{13}	 &\multirow{2}{*}{79.8}	&VBF+$VH$	&$1.23\pm0.26$&\multirow{2}{*}{\cite{Aad:2019mbh}}\\
    &	&	&	&$t\bar th+th$	&$0.79^{+0.60}_{-0.59}$	&\\\cline{2-7}
 	&\multirow{2}{*}{CMS}	&\multirow{2}{*}{7+8+13}	&\multirow{2}{*}{41.3}	&VH (0-2$\ell$, 2 b-tags+jets)	&$1.01\pm0.22$&\multirow{2}{*}{\cite{Sirunyan:2018kst}}\\
 	&	&	&	&all prod.	&$1.04\pm0.2$&\\ \cline{2-7}
 	&\multirow{2}{*}{ATLAS+CMS}	&\multirow{2}{*}{7+8+13}	&\multirow{2}{*}{}	&VH, \pc	 &$0.98\pm0.15$&\multirow{2}{*}{\cite{Sirunyan:2018kst,Aaboud:2018zhk}}\\
 	&	&	&	&all prod., \pc	&$1.02\pm0.14$&\\
 	\hline\hline
\end{tabular} 	
\label{tab:muexp}
\end{center}
\end{table}

\begin{table}[h]
 \begin{center}
 \caption{Collider limits on the Higgs signal strengths involving the  $t$. 
 In our private combinations, the uncertainties are approximated as symmetric. The $h\to\gamma\gamma$ decay does not lead to competitive bounds on $T_R^t,\,T_I^t$.
 }
 \label{tab:top}
  \begin{tabular}{||c||c|c|c|c|c||c||}
   \hline \hline
   channel &experiment	&$\sqrt{s}/\tev$ &$\mathscr{L}/\ifb$ &comment 	&$\mu$	&Ref\\ \hline \hline
\multirow{9}{*}{ggF}	
 	&\multirow{3}{*}{ATLAS}	&\multirow{3}{*}{13}	&\multirow{3}{*}{$\leq 79.8$}	&$H\to\gamma\gamma$	 &$0.96\pm0.14$&\multirow{3}{*}{\cite{Aad:2019mbh}}\\
 	&	&	&	&$H\to\tau\tau$	&$0.96^{+0.59}_{-0.52}$&\\
 	&	&	&	&all decays, fixed to SM	&$1.04\pm0.09$&\\ \cline{2-7}
 	&\multirow{3}{*}{CMS}	&\multirow{3}{*}{13}	&\multirow{3}{*}{$35.9$}	&$H\to\gamma\gamma$	& $1.16^{+0.30}_{-0.25}$&\multirow{3}{*}{\cite{Sirunyan:2018koj}}\\
 	&	&	&	&$H\to\tau\tau$	&$1.05^{+0.75}_{-0.67}$ &\\
 	&	&	&	&all decays, fixed to SM	&$1.22^{+0.20}_{-0.18}$&\\ \cline{2-7}
 	&\multirow{3}{*}{ATLAS+CMS}	&\multirow{3}{*}{13}	&\multirow{3}{*}{}	&$H\to\gamma\gamma$, \pc	 &$1.00 \pm 0.12$&\multirow{3}{*}{\cite{Aad:2019mbh,Sirunyan:2018koj}}\\
 	&	&	&	&$H\to\tau\tau$, \pc	&$0.99\pm0.44$&\\
 	&	&	&	&all decays SM, \pc	&$1.07\pm0.08$&\\ \hline
\multirow{9}{*}{$\bar tth+ th$}	
 	&\multirow{4}{*}{ATLAS}	&\multirow{4}{*}{13}	&\multirow{4}{*}{$\leq 79.8$}	&$H\to\gamma\gamma$	 &$1.1^{+0.41}_{-0.35}$ &\multirow{4}{*}{\cite{Aad:2019mbh}}\\
 	&	&	&	&$H\to\tau\tau$	&$1.38^{+1.13}_{-0.96}$&\\
    &	&	&	&$H\to\bar b b$	&$0.79^{+0.60}_{-0.59}$&\\
 	&	&	&	&all decays, fixed to SM	&$1.21^{+0.26}_{-0.24}$&\\ \cline{2-7}
 	&\multirow{4}{*}{CMS}	&\multirow{4}{*}{13}	&\multirow{4}{*}{$35.9$}	&$H\to\gamma\gamma$	& $2.18^{+1.25}_{-1.06}$&\multirow{4}{*}{\cite{Sirunyan:2018koj}}\\
 	&	&	&	&$H\to\tau\tau$	&$0.23^{+1.46}_{-1.24}$ &\\
 	&	&	&	&$H\to\bar b b$	&$0.91^{+0.64}_{-0.61}$ &\\
 	&	&	&	&all decays, fixed to SM	&$1.18^{+0.43}_{-0.38}$&\\ \cline{2-7}
 	&\multirow{4}{*}{ATLAS+CMS}	&\multirow{4}{*}{13}	&\multirow{4}{*}{}	&$H\to\gamma\gamma$, \pc	 &$1.21\pm0.36$&\multirow{4}{*}{\cite{Aad:2019mbh,Sirunyan:2018koj}}\\
 	&	&	&	&$H\to\tau\tau$, \pc	&$0.95\pm 0.83$&\\
 	&	&	&	&$H\to\bar bb$, \pc	&$0.87\pm0.43$&\\
 	&	&	&	&all decays SM, \pc	&$1.20\pm0.21$&\\ \hline
\multirow{2}{*}{ggF+$\bar tth+\bar t H$}	&\multirow{2}{*}{ATLAS+CMS}	&\multirow{2}{*}{13}		&	&all decays SM, \pc	&$1.09\pm0.08$&\multirow{2}{*}{\cite{Aad:2019mbh,Sirunyan:2018koj}}\\
&	&	&	&$h\to VV$, \pc	&$1.08\pm0.08$&\\ \hline
 	    \multirow{4}{*}{$h\to \gamma\gamma$}	
 	&ATLAS	&\multirow{4}{*}{13}		&$138$	&\multirow{2}{*}{$tth$}	 &$1.38^{+0.41}_{-0.36}$&\cite{Keller:2680951,ATLAS:2019aqa}\\
 	&CMS	&	&$77.4$	&	&$1.7^{+0.6}_{-0.5}$&\cite{Keller:2680951,CMS:2018rbc}\\
 	&ATLAS	&	&$\leq 79.8$	&\pc~of all prod.	&$1.02\pm0.12$&\cite{Aad:2019mbh}\\
 	&CMS	&	&$35.9$	&SM ratio of production modes&$1.2^{+0.25}_{-0.20}$&\cite{Sirunyan:2018koj}\\	 
 	\hline\hline
  \end{tabular}
  \label{tab:muexpt}
 \end{center}
\end{table}

\clearpage
\subsection{Analytical bounds for $T_R^f=0$}
\label{app:TR=0}
When we consider simultaneous modifications of two Yukawa couplings in Figs.~\ref{fig:btau_TR0} - \ref{fig:tb_TR0}, we take as an example the case of $T_R^f=0$. Here we provide the conditions on the relations between $\br_f^\sm$ and the experimental bounds on $\mu_I^F$ under which a bound on $T_I^{f_1}$, $T_I^{f_2}$ arises in the two possible cases presented in Table ~\ref{tab:categoriesf1f2}.

\paragraph{Case A} If either the production or the decay is not modified with respect to the SM (namely VBF/$Vh$ production or the decay $h\to VV$), the signal strength is given by Eq.~(\ref{eq:prodSM_decf1f2}). Depending on whether the measured signal rate is $\mu_{f_1}>$ or $<1$, the resulting exclusion contour has an asymptote at
\begin{align}
 T_{I}^{f_1,{\rm asymp}} &= \pm\sqrt{\frac{1-\mu_{f_1}}{\mu_{f_1}(1+8\br^\sm_{f_1})-9}}\,,
 ~~{\rm if}~\frac{9}{1+8\br^\sm_{f_1}}>\mu_{f_1}>1\,,\label{eq:TI1a}\\
 T_{I}^{f_2,{\rm asymp}} &= \pm\sqrt{\frac{1-\mu_{f_1}}{\mu_{f_1}(1+8\br^\sm_{f_2})-1}}\,,
 ~~{\rm if}~\frac{1}{1+8\br^\sm_{f_2}}<\mu_{f_1}<1\,.\label{eq:TI2a}
\end{align}
For example, the $2\sigma$ lower bound on $\mu_{{\rm VBF}+Vh}^{\tau\tau}$ of $0.57$ is smaller than $1/(1+8\br_g^\sm)=0.59$, see Eq.~(\ref{eq:TI2a}). Therefore, this lower bound does not imply a limit on $T_I^\tau$, $T_I^t$ in Fig.~\ref{fig:ttau_TR0}, but a slight experimental improvement will give rise to a limit.
A measurement of $\mu_{f_1}=1$ results in an X-shaped contour crossing through the origin
$\left(T_I^{f_1}, T_I^{f_2}\right)=(0,0)$.

\paragraph{Case B}
The production via the coupling $\lambda_{f_1}$ (hence $\lambda_t$) and the decay $h\to f_2\bar f_2$,
while the total width is modified by both $\lambda_{f_1}$ and $\lambda_{f_2}$,
are described by the signal strength in Eq.~(\ref{eq:prodf1_decf1f2}).
It results in a bound if
\begin{equation}
\label{eq:muf1f2Condition}
 1\leq \mu_{f_1}^{f_2} \leq \frac{9}{1+8 \br_{f_1}^\sm},\frac{9}{1+8 \br_{f_2}^\sm}\,.
\end{equation}
For $\mu_{f_1}^{f_2}<1$ there is no real solution.
Consequently, the lower bound on $\mu_{\rm ggF}^{\tau\tau}$ does not yield a limit in the $T_I^\tau$-$T_I^t$ plane.
In case $\mu_{f_1}^{f_2}=1$ is measured, the only solution is the trivial one, namely the SM prediction of $T_I^\tau=T_I^t=0$, or the physically impossible case of
$\br_g^\sm=\br_\tau^\sm=1$.

\section{Benchmark values for baryogenesis calculations}\label{sec: benchmark}
In this appendix we present benchmark values for various parameters required to reproduce our results. Further expressions and details can be found in \cite{flnv}.
We take the nucleation temperature to be $ T_N = 88 $ GeV. At this temperature, the gauge couplings and Higgs VEV are \cite{deVries:2017ncy}
\begin{align}
g' &= 0.36 \,, & g &= 0.65 \,, & g_s &= 1.23 \,, & v_N &= 152 \, \text{GeV} \,.
\end{align}
The entropy density, written in terms of the temperature and the entropy degrees of freedom $ g^* $, is given by \cite{deVries:2017ncy}
\begin{align}
s = \frac{2\pi ^2}{45}g^* T_N^3 \,, \ \ \ g^* = 106.75 .
\end{align}
The bubble wall velocity and width are taken from \cite{deVries:2018tgs}, with values
\begin{align*}
v_w = 0.05 \,, \ \ \ L_w = 0.11 \, \text{GeV}^{-1} \,,
\end{align*}
respectively. We apply the VEV-insertion approximation as well as the diffusion approximation. The diffusion coefficients are approximately given by \cite{Joyce_1996,Cline_2000}
\begin{align}\label{eq:diffusionC}
D_{l_R} &= 380/T \,, & D_{l_L} &= 100/T \,, & D_u &= D_d = D_q = 6/T \,, & D_h = 100/T \,. 
\end{align}
The thermal widths are \cite{Elmfors_1999}
\begin{align}
\Gamma _{\text{lepton}} \approx 0.002 T \,, \ \ \ \Gamma _{\text{quark.}} \approx 0.16 T \,.
\end{align}
Finally, the sphaleron rates are given by \cite{Bodeker:1999gx,Moore:2010jd}
\begin{align}
\Gamma _{\text{ws}}(z) = 120\alpha_w^5 T_N \Theta(z<0) \,, \ \ \ \Gamma _{\text{ss}} = 14\alpha _s^4 T_N \,.
\end{align}

\bibliography{EWBGdim6}
\markboth{}{}
\end{document}